\documentclass[a4paper,11pt]{article}
\pdfoutput=1 
\usepackage{jcappub}
\usepackage[T1]{fontenc}
\usepackage{mathrsfs}
\usepackage{hyperref}
\usepackage{bm}
\usepackage[caption=false]{subfig}
\usepackage[normalem]{ulem}
\usepackage{mathtools}
\usepackage{array}
\newcolumntype{P}[1]{>{\centering\arraybackslash}p{#1}}
\newcolumntype{M}[1]{>{\centering\arraybackslash}m{#1}}

\DeclareMathOperator\erf{erf}
\DeclareMathOperator{\csch}{csch}

\def\nue{\mathrel{{\nu_e}}}

\def\d{\delta}

\def\t13{\mathrel{{\theta_{13}}}}
\def\y12{\mathrel{{\tan^2 \theta_{12}}}}
\def\c2{\mathrel{{\chi^2 }}}

\def\msun{\mathrel{{M_{\odot} }}}

\newcommand{\ans}{anisotropy}
\newcommand{\gw}{GW}
\newcommand{\n}{neutrino}
\newcommand{\ns}{neutrinos}
\newcommand{\sn}{supernova}
\newcommand{\sne}{supernovae}
\newcommand{\mms}{multi-messenger}
\newcommand{\mm}{memory}

\newcommand{\be}{\begin{equation}}
\newcommand{\ee}{\end{equation}}
\newcommand{\ba}{\begin{eqnarray}}
\newcommand{\ea}{\end{eqnarray}}

\newcommand{\half}{\frac{1}{2}}
\newcommand{\pd}{{\partial}}
\newcommand{\np}{{n^{\prime}}}

\title{\boldmath 
The neutrino gravitational memory from a core collapse supernova: phenomenology and physics potential
}

\author[1]{Mainak Mukhopadhyay,\note{Corresponding author.}}
\author{Carlos Cardona}
\author{and Cecilia Lunardini}

\affiliation{Department of Physics, Arizona State University, Tempe, AZ 85287, USA.}

\emailAdd{mmukhop2@asu.edu}
\emailAdd{ccardon6@asu.edu}
\emailAdd{Cecilia.Lunardini@asu.edu}

\abstract{
General Relativity predicts that the passage of matter or radiation from an asymmetrically-emitting source should cause a permanent change in the local space-time metric. This phenomenon, called the \emph{gravitational memory effect}, has never been observed, however supernova neutrinos have long been considered a promising avenue for its detection in the future. With the advent of deci-Hertz gravitational wave interferometers, observing the  supernova neutrino memory will be possible, with important implications for multimessenger astronomy and for tests of gravity. In this work, we develop a phenomenological (analytical) toy model for the supernova neutrino memory effect, which is overall consistent with the results of numerical simulations. This description is then generalized to several case studies of interest.  We find that, for a galactic supernova, the dimensionless strain, $h(t)$, is of order $\sim 10^{-22} - 10^{-21}$, and develops over a typical time scale that varies between $\sim 0.1 - 10$ s, depending on the time-evolution of the anisotropy of the neutrino emission. The characteristic strain, $h_c(f)$, has a maximum at a frequency $f_{max} \sim {\mathcal O}(10^{-1}) - {\mathcal O}(1)$ Hz. The detailed features of the time- and frequency-structure of the memory strain will inform us of the matter dynamics near the collapsed core, and allow to distinguish between different stellar collapse scenarios.  Next generation gravitational wave detectors like DECIGO and BBO will be sensitive to the neutrino memory effect for supernovae at typical galactic distances and beyond; with Ultimate DECIGO exceeding a detectability distance of 10 Mpc. 
}

\begin{document}
\maketitle
\flushbottom

\section{Introduction}
\label{sec:intro}

Neutrino- and gravitational wave-astronomy are emerging players in the new field of \mms\ astronomy. They both have the potential to investigate ``dark'' phenomena like the core collapse of a massive star and its possible implosion into a black hole, and the process of inspiral and merger of a binary systems involving at least one neutron star. After the LIGO-Virgo observation of binary mergers \cite{Abbott:2016blz, TheLIGOScientific:2017qsa, Abbott:2020uma, Abbott:2020tfl}, such exploration is already a reality for gravitational waves (\gw ), and a similar level of steady progress might be achieved in the next decades with the next generation of low background \n\ observatories reaching up to a Megaton mass. While \n\ and \gw\ physics are still mostly developing separately, their potential as complementary probes of the same astrophysical phenomena has recently been recognized, and dedicated, interdisciplinary research has begun. 


Surprisingly,  so far only limited attention has been paid to the most \emph{direct} connection between \ns\ and \gw: the \emph{gravitational memory} caused by (anisotropic) \n\ emission. The essence of this effect has been known since the 1970's \cite{Zeldovich:1974gvh, 1987Natur.327..123B}: anisotropic \n\ emission, for example by a core collapse \sn, would cause a non-oscillatory, \emph{permanent} strain in the spacetime metric that would in principle be visible at a powerful \gw\ detector \cite{Epstein:1978dv,Turner:1978jj}. The theory of the \mm\ effect is well established, having been developed at the formal level for a generic emitter of radiation and matter \cite{Sago:2004pn, Suwa_2009,Favata:2010zu}. Early applications to a core collapse supernova were developed as well, analytically and numerically \cite{Epstein:1978dv,Turner:1978jj,Burrows:1995bb,janka_muller}. 
Results showed that, in a \gw\ detector, the  neutrino-induced \mm\ from a galactic \sn\ would appear as a signal with typical frequency of $0.1-10$ Hz and  (dimensionless) strain of $\sim 10^{-22}-10^{-20}$, which is well below the sensitivity of LIGO and its immediate successors. Long considered unobservable, the \mm\ has thus largely remained a textbook-case curiosity.

This situation is about to change with the third generation of \gw\ detectors, especially those designed to explore the  Deci-Hz frontier \cite{Seto:2001qf,Yagi_2011,Luo_2016,PhysRevD.94.104022,Sato_2017,amaroseoane2017laser, Guo:2018npi}, namely the region centered at frequency $f\sim 0.1$ Hz. Ambitious projects like the DECi-hertz Gravitational-wave Observatory (DECIGO) \cite{Seto:2001qf,Yagi_2011} and the Big Bang Observer (BBO) \cite{Yagi_2011} will reach a sensitivity of $\sim 10^{-24}$ in strain, and therefore will be able to observe the \sn\ \n\ \mm. An experimental test of the \mm, with its distinctive hereditary nature, would be an important confirmation of general relativity. 
The new observational prospects have stimulated several modern studies of the \sn\ \n\ memory, based on state-of-the art hydrodynamic simulations in two and three dimensions \cite{Burrows:1995bb,Kotake_2007,Kotake:2009rr,Muller:2011yi,Yakunin:2015wra,Vartanyan:2020nmt}, where the detailed time structure of the neutrino luminosity and of the anistropy parameter are modeled. Due to computational cost, simulations have been conducted for isolated examples of progenitor star, and reproduce only part of the \mm\ evolution, up to about $1$ s after the core bounce. 

In the light  of the recent advancements on modeling the \mm\ effect numerically, the time is now mature for the development of phenomenological studies, for the benefit of the broader community, with the goal of assessing the detectability and physics potential of the \sn\ \n\ \mm\ effect.  These studies will necessarily require a parametric description that captures the essential features of the \mm\ over the entire $\sim 10$ s of duration of the \n\ burst, and can be applied to wide variety of phenomenological scenarios, corresponding to different stellar progenitors, different outcomes of the collapse (successful explosion or implosion into a black hole), etc.  Such description can be useful as a foundation for more advanced phenomenological studies, and as a template to simulate the response of a \gw\ detector to a \mm\ signal. 

The present paper is a first effort in this direction. We develop a phenomenological model of the \n-induced memory strain both in time and frequency domain. Our model is sufficiently realistic -- because it is based on realistic (although simplified) assumptions, motivated by numerical results for the \mm\ and at the same time is sufficiently simple to be used widely. We apply it to a number of plausible core collapse scenarios, and discuss the physics potential of a joint detection of a \n\ burst and a \mm\ signal from a galactic \sn. Our study extends and complements previous analytical description of the \mm, which were developed for long gamma-ray bursts (GRBs)~\cite{Sago:2004pn} where the memory from a single jet and a unified model of the GRB were considered along with the angular dependence of the wave-form, supermassive stars~\cite{Li:2017mfz} which estimated the memory strain magnitude from supermassive stars ($\sim 5 \times 10^4 M_\odot$) and discussed the prospects of their detection and hypernovae~\cite{Suwa_2009} using spherically symmetric and oblate-spheroidal accretion discs and constant \n\ anisotropy parameter. 
 
The paper is structured as follows. In Sec. \ref{sec:formalism} the formalism describing the \mm\ signal is summarized, and general upper bounds are derived. Our model is introduced in Sec. \ref{sec:pheno}; then in Sec. \ref{sec:results} we present applications and discuss the detectability of the \sn\ \n\ \mm\ at future Deci-Hz interferometers. A discussion section, Sec. \ref{sec:discussion}, concludes the paper, followed by technical appendices.  
 
\section{Formalism}
\label{sec:formalism}
In this section we summarize the formalism that describes the \mm\ strain due the anisotropic emission of radiation  (massless particles) by a generic source, and specialize it to neutrinos from a core collapse supernova. 

\subsection{Memory wave form}

For completeness, here we review the theoretical framework of the \mm,
following closely some classic papers on the subject  \cite{Weinberg:1972kfs,Misner:1974qy, Epstein:1978dv}. For brevity, certain derivations are omitted;   we refer the reader to appendix \ref{sec:appendix A} for those. 

The starting point is  Einstein's field equation,
\be
\label{eqn:einseqn}
R_{\mu \nu} - \half R g_{\mu \nu} = -8 \pi G T_{\mu \nu}\,,
\ee
where, $R_{\mu \nu}$ is the Ricci tensor, the Ricci scalar $R = 8 \pi G T^\mu_\mu$, $g_{\mu \nu}$ is the metric, $G$ is the Newton's universal gravitational constant and $T_{\mu \nu}$ is the stress-energy tensor\footnote{The detailed expressions for each are given in Appendix~\ref{sec:appendix A}}.

\begin{figure}[htb]
\begin{center}
\includegraphics[width=0.65\textwidth,angle=0]{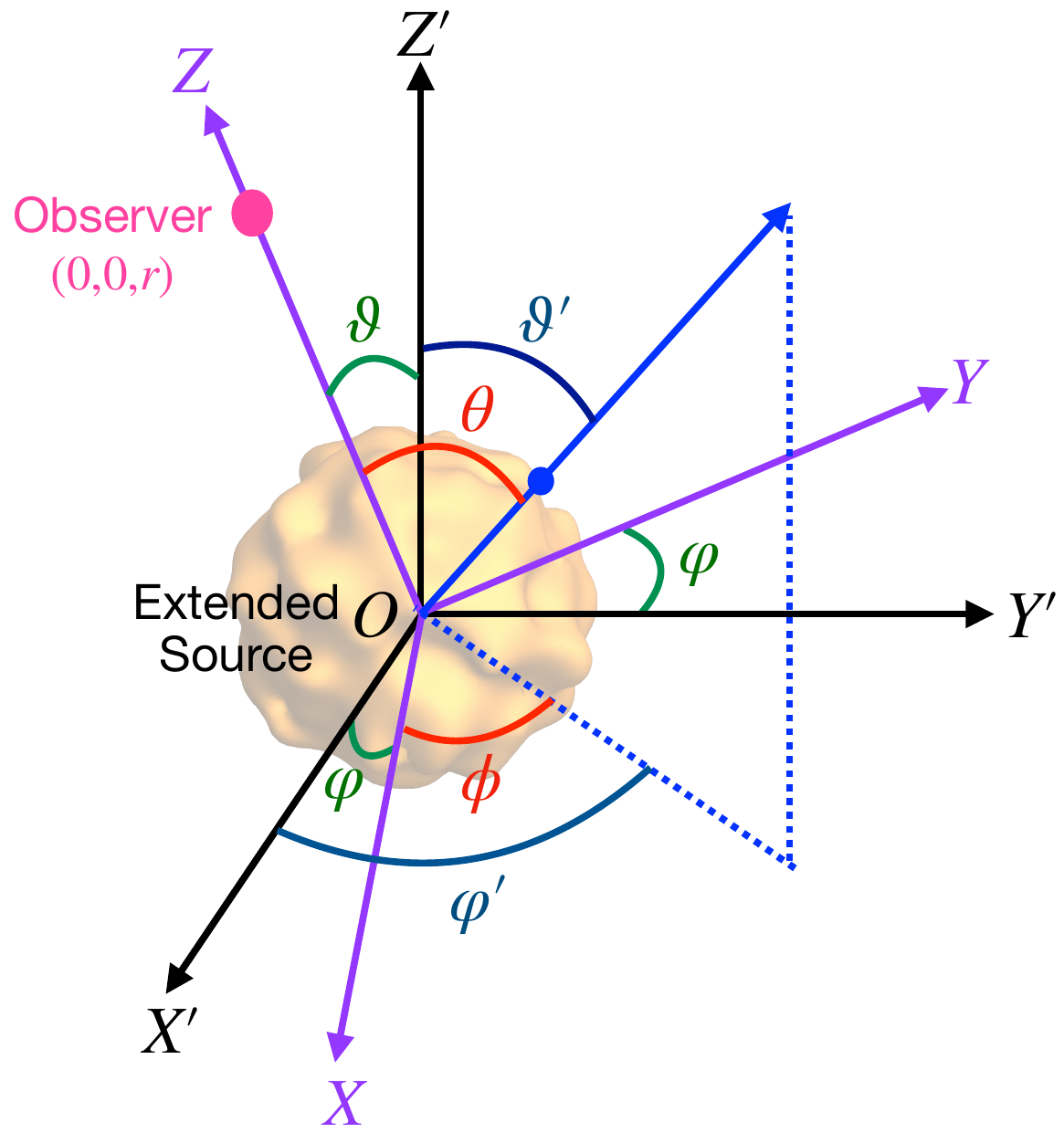}
\caption{\label{fig:setup1}Setup of the coordinate systems - $(XYZ)$ (Purple): Coordinate system for observer , $(X^{\prime}Y^{\prime}Z^{\prime})$ (Black): Coordinate system for source (source is treated as an extended source, centered at the origin; denoted by yellow blob). The blue dot is a point on the surface of the extended source and the corresponding position vector is shown as a blue arrow. $(\vartheta^{\prime},\varphi^{\prime})$ (Dark Blue): Radiation direction in the source frame $(X^{\prime}Y^{\prime}Z^{\prime})$; $(\theta, \phi)$ (Red): Radiation direction in the observer's frame $(XYZ)$, $(\vartheta, \varphi)$ (Dark Green): Orientation of observer's frame $(XYZ)$ with respect to the source frame $(X^{\prime}Y^{\prime}Z^{\prime})$. Observer (Pink) is located along the $Z$-axis at a distance, $r$ $(0,0,r)$.
}
\end{center}
\end{figure}

Here it suffices to work in the weak-field approximation, where the metric is nearly flat, with small perturbations $h_{\mu \nu}$: 
\be
\label{metric}
g_{\mu \nu} = \eta_{\mu \nu} + h_{\mu \nu}.
\ee
In this approximation, the field equations ~\eqref{eqn:einseqn}  are still invariant under coordinate transformations that preserve the weak-field condition. We can use this gauge freedom to choose a convenient gauge: $g^{\mu \nu} \Gamma^\lambda_{\mu \nu} = 0$ ($\Gamma^\lambda_{\mu \nu}$ is the Christoffel symbol defined in Appendix~\ref{sec:appendix A}). By keeping only terms up to first order in $h_{\mu \nu}$, from \eqref{eqn:einseqn} we get the following wave equation in flat space for the field perturbation,
\be
\label{eqn:mainfieldeqn}
\Box^2 h_{\mu \nu} = -16 \pi G S_{\mu \nu}\,,
\ee
where the effective stress-energy tensor (in presence of matter) is 
$S_{\mu \nu} = T_{\mu \nu} - \half \eta_{\mu \nu}T^\lambda_\lambda$.
One can solve the wave equation by using the retarded Green's function corresponding to the D'Alembert operator in four-space time dimensions, so to obtain the expression:
\be
\label{eqn:greensol}
h_{\mu \nu} = 4G \int d^3 \vec{x}^{\,'} \Big( \frac{S_{\mu \nu}(\vec{x}^{\,'},t-|\vec{x}-\vec{x}^{\,'}|)}{|\vec{x}-\vec{x}^{\,'}|} \Big)\,.
\ee
The gauge choice leading us to this solution does not fix completely all the gauge freedom and an additional constraint should be imposed to leave only the physical degrees of freedom. This is done by projecting the source tensor $S_{\mu \nu}$ into its transverse-traceless (TT) components (see for example \cite{Misner:1974qy}). Doing this and without loss of generality, we will use the following very well known ansatz for the source term proposed in \cite{Epstein:1978dv},  
\be
\label{eqn:source}
S^{ij}(t,x) = \frac{( n^i n^j)_{TT}}{r^{2}}\int_{-\infty}^{\infty} \sigma(t')f(\Omega',t')\delta(t-t'-r)dt',
\ee
where, $\vec{n} = \vec{x}/r$,  $r=|\vec{x}|$ and the sub index (TT) denotes the transverse-traceless component. This source term represents the effect of emitted  radiation from the source origin at $x=0$ \footnote{Due to the conservation of the stress-energy tensor, we only need to consider the spatial index of the tensor}.
Here $\sigma(t)$ denotes the rate of energy loss, and $f(\Omega',t')$ is the angular distribution of emission, where the argument $\Omega'$  is a shorthand notation indicating the dependence on the angles $\vartheta^{\prime}$ and $\varphi^{\prime}$.  $d\Omega'$ is the differential solid angle, $d\Omega'=\sin(\vartheta^{\prime})d\vartheta^{\prime}d\varphi^{\prime} $ (see Fig.~\ref{fig:setup1}), so that $f (\Omega',t')\geq 0$ and $\int f(\Omega',t')d\Omega' = 1$.
After substituting the ansatz \eqref{eqn:source} into the wave form \eqref{eqn:greensol}, and integrating, we obtain the following expression for the wave form:
\be
\label{eqn:httform1}
h^{ij}_{TT}(t,x) = 4 G \int_{-\infty}^{t-r} \int_{4 \pi} \frac{(n^i n^j)_{TT} f(\Omega^{\prime},t^{\prime}) \sigma(t^{\prime})}{t-t^{\prime}-r \cos{\theta}} d \Omega^{\prime} dt^{\prime}\,,
\ee
where $\theta$ is the angle between the observer position and the radiation source point. Following \cite{janka_muller}, we assume that the observer is situated at a distance $r=|x|\rightarrow \infty$ from the source. 
The  radiation that reaches the observer at a time $t$ was actually emitted at time, $t^{\prime} = t-r/c$, physically representing a case where the neutrino pulse itself causes a gravitational wave signal.
We can now rewrite~\eqref{eqn:httform1} in this approximation  as,
\be
\label{eqn:httform2a}
h^{ij}_{TT}(t,x) = \frac{4G}{r c^4} \int_{-\infty}^{t-r/c} dt^{\prime} \int_{4 \pi} \frac{(n^i n^j)_{TT}}{1 - \cos{\theta}} \frac{dL_\nu (\Omega^{\prime},t^{\prime})}{d \Omega^{\prime}} d \Omega^{\prime},
\ee
where, $f(\Omega^{\prime},t^{\prime}) \sigma(t^{\prime}) = \frac{dL_\nu (\Omega^{\prime},t^{\prime})}{d \Omega^{\prime}}$, which is the direction dependent neutrino luminosity. 

Fig.~\ref{fig:setup1} shows the orientation of the coordinate axes for the observer and the source. The different angles involved are also shown. The wave $h^{ij}_{TT}(t,x)$ can be either `$+$' or `$\times$' polarized. We denote the `$+$' polarization as, $h^{xx}_{TT} = -h^{yy}_{TT} = -h^{+}_{TT}$. With this in mind, we now need to compute the different pieces of~\eqref{eqn:httform2a}. One obtains (see  Appendix~\ref{appsec:nxnxtt}) $(n^x n^x)_{TT} = \frac{1}{2} (1 - \cos^2 {\theta})(2 \cos^2{\phi} - 1) = \frac{1}{2}(1 - \cos^2 {\theta})\cos{2 \phi}$. Substituting this in Eq.~\eqref{eqn:httform2a} gives,
\be
\label{eqn:transamplitude}
h^{xx}_{TT} = \frac{2G}{r c^4} \int_{-\infty}^{t-r/c} dt^{\prime} \int_{4 \pi} (1+\cos{\theta})\cos{2\phi} \frac{dL_\nu (\Omega^{\prime},t^{\prime})}{d \Omega^{\prime}} d \Omega^{\prime}.
\ee

The `$\times$' polarization, $h^{xy}_{TT} = h^{\times}_{TT}$ can be found by simply replacing $\cos{2 \phi}$ by $\sin{2 \phi}$.
One can isolate the angular dependence by defining the \emph{anisotropy parameter} $\alpha(t)$ as,
\be
\label{eqn:anisotropy}
\alpha(t) = \frac{1}{L_{\nu}(t)} \int_{4 \pi} d \Omega^{\prime}\ \Psi (\vartheta^{\prime}, \varphi^{\prime})\ \frac{dL_{\nu} (\Omega^{\prime},t)}{d \Omega^{\prime}}\,,
\ee 
where $\Psi (\vartheta^{\prime}, \varphi^{\prime})$ is an angle-dependent function that depends solely on the location of the observer with respect to the source, i.e., $\theta$ and $\phi$ appearing in Eq.~\eqref{eqn:transamplitude} are expressed in terms of $\vartheta^{\prime}$ and $\varphi^{\prime}$ based on the observer's location with respect to the source (see~\cite{janka_muller,Kotake:2009rr} for example and details.) 

This enables us to write~\eqref{eqn:transamplitude} in the following convenient form,
\be
\label{eqn:mainampeq}
h^{xx}_{TT} = h(t) = \frac{2G}{r c^4} \int_{-\infty}^{t-r/c} dt^{\prime} L_{\nu}(t^{\prime}) \alpha(t^{\prime})\,.
\ee

The anisotropy parameter plays a very significant role in determining the amplitude of the gravitational wave strain $h(t)$. It is mainly governed by the complex dynamics of the source. 
We will discuss its role in more detail in later sections. 
It is important to note here that if an ideal gravitational wave detector has two freely falling masses which have a vectorial separation of $l_k$, the gravitational wave strain changes their separation by $\delta l_j$ where,
\be
\delta l_j = \frac{1}{2} h_{jk}^{TT} l^k\,.
\ee
Of course in Eq.~\eqref{eqn:mainampeq}, we have just considered the strain in the x-directon. It may also be useful to express the gravitational wave strain $h(t)$ in frequency space,
\be
\Tilde{h} (f) = \int_{-\infty}^{\infty} h (t)\  e^{2 \pi i\, f t}\ dt\,,
\label{eqn:fourier}
\ee
where, $\Tilde{h} (f)$ is the Fourier transform of $h (t)$. Finally, we define the characteristic strain $h_c (f)$~\cite{Sago:2004pn,Li:2017mfz} as,
\be
 h_c(f) = 2f|\Tilde{h}(f)|\,.
\label{eqn:hchar}
\ee
This is a dimensionless quantity (the $f$ appearing above makes it dimensionless), which helps in computing the signal to noise ratio (SNR) for a given gravitational wave detector and compare the signal to the sensitivity curve of the detector to predict the prospects of detection of the signal using the given detector.

\subsection{General properties of the neutrino memory signal}
\label{subsec:genprop}

In this section we discuss properties of the \mm\ signal, in the time and frequency domain, that stem directly from its expression as an integral over time, Eq. \eqref{eqn:mainampeq}, and therefore have general validity.  

\subsubsection{Time domain: evolution and upper bound } 

\begin{figure}
\begin{center}
\includegraphics[width=0.6\textwidth]{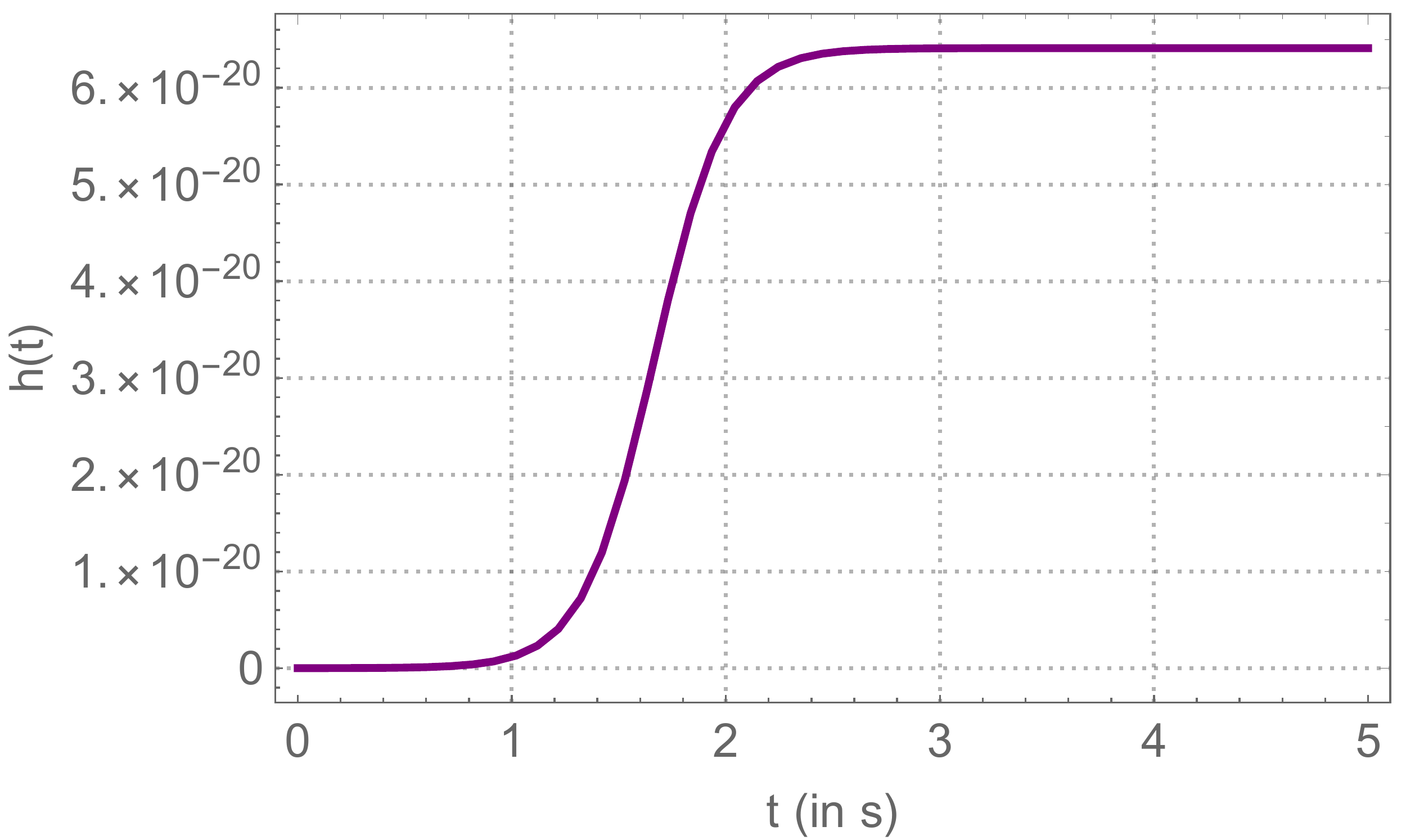}
  \caption{ \label{fig:ameba} Sketch of a typical gravitational wave memory strain profile, $h(t)$. 
  }  
\end{center}
\end{figure}

Considering the finite duration ($\Delta t \sim 10$ s) of the \n\ burst, from \eqref{eqn:mainampeq} we expect  the metric perturbation $h(t)$, to transition from an asymptotic value $h=0$ at earlier times ($t\rightarrow -\infty$) to a different asymptotic value $h=\Delta h$ at later times ($t\rightarrow +\infty$), as sketched in Fig.~\ref{fig:ameba}. The characteristic rise time must be $\delta t \lesssim \Delta t$, depending on the time dependence of $\alpha(t)$.  
In physical terms, the gravitational memory accumulates from the arrival of the first \ns\ to Earth until the \n\ burst  has passed completely. 

We can place a conservative upper limit on $h(t)$ from the following inequality: 
\begin{equation}
\label{eqn:uppert}
|h(t)| 
\leq  \frac{2G}{r c^4} \int_{-\infty}^{\infty}L_{\nu}(t) |\alpha(t)| dt \leq 
 \frac{2G}{r c^4} |\alpha|_{max} E_{tot}~. 
\end{equation}
Here we accounted for the possibility that $\alpha(t)$ may be negative and change sign (see Sec. \ref{subsec:template}), and  $|\alpha|_{max}$ is the maximum of its value (in modulus). 
$E_{tot}=\int_{-\infty}^{\infty}L_{\nu}(t) dt \simeq 3 \times 10^{53}~{\rm ergs}$ is the total energy emitted by neutrinos.
Numerically, Eq.  (\ref{eqn:uppert}) gives: 
\be
\label{eqn:uppernum}
|h(t)|  \leq 6.41~  10^{-20} \left(\frac{|\alpha|_{max}}{0.04} \right) \left(\frac{E_{tot}}{3 ~10^{53}~{\rm ergs}} \right)  \left(\frac{r}{ 10~  {\rm kpc}} \right)^{-1} ~.   
\ee
The same bound holds for $|\Delta h|$, as one can easily verify.

\subsubsection{Frequency domain: limiting cases}
\label{subsec:ZFLHFL}

Let us now discuss the main features of $h_c(f)$ (Eq. \ref{eqn:hchar}). We expect it to be dominated by frequencies of the order of $f_c \sim 1/2 \pi \delta t \gtrsim 1/2 \pi \Delta t \sim 0.02$ Hz.

In agreement with previous literature~\cite{Turner:1978jj,Favata:2011qi}, in the \emph{zero frequency limit} (ZFL),  $f\ll f_c$, $h_c(f)$ tends to a constant value. This can be proven by observing that: 
\ba
\label{eq:dotfourier}
\lim_{f\to 0} h_c(f)= \lim_{f\to 0}| 2 f \tilde h| = \lim_{f\to 0} \left|  \frac{i}{\pi} \tilde{\dot{h}} \right | ~,
\ea
where we used the well known property relating the Fourier Transform of a function to the Fourier Transform of its derivative ($\tilde{\dot{g}}(f)=-2\pi i\, f \tilde{g}(f)$) (see appendix \ref{appendix_fourier_derivative} for the validity of this property in the case at hand). 
Now, we note that
\ba
\lim_{f\to 0}\tilde{\dot{h}}&=&\lim_{f\to 0}\int_{-\infty}^{\infty}\dot{h}(t)e^{2\pi i\, f\, t} dt
=\int_{-\infty}^{\infty}\dot{h}(t) dt=\Delta h\,,
\ea
leading immediately to the result
\be 
\label{eq:zfl}
\lim_{f\to 0} h_{c} = \frac{|\Delta h|}{\pi}~,
\ee
which does not depend on $f$. 
Therefore, we expect a flat behaviour of $h_{c}$ at low frequencies, characterized by the strength of the metric change $\Delta h$ \footnote{The ZFL provides a good estimate of the wave strain when the time scale of the burst is much longer than the neutrino emission process \cite{Turner:1978jj}. Additionally, at the ZFL the neutrino quantum production can be computed classically. }. 

When combined with Eq. (\ref{eqn:uppert}), Eq. (\ref{eq:zfl}) gives an upper limit: 
\be 
\label{eq:zfllim}
\lim_{f\to 0} h_{c} \leq  \frac{2G}{\pi r c^4} |\alpha|_{max} E_{tot} \simeq  2.0~  10^{-20} \left(\frac{|\alpha|_{max}}{0.04} \right) \left(\frac{E_{tot}}{3 ~10^{53}~{\rm ergs}} \right)  \left(\frac{r}{ 10~  {\rm kpc}} \right)^{-1} ~. 
\ee

The latter bound can be shown to be valid at all frequencies. Indeed, consider that, from Eq. (\ref{eqn:transamplitude}), we can write 
\be
\label{dothexpr}
\dot{h}(t)= \frac{2G}{r c^4} L_{\nu}(t)\alpha(t)\,. 
\ee
Therefore 
\be
\label{eq:allfreqlim}
|\tilde{\dot{h}}|= \frac{2G}{r c^4} \left|\int_{-\infty}^{\infty}L_{\nu}(t)\alpha(t)e^{2\pi i\, f\, t} dt\right|\leq  \frac{2G}{r c^4} \int_{-\infty}^{\infty}L_{\nu}(t)\,|\alpha(t)| dt\,.
\ee
This result can then be combined with Eqs. (\ref{eqn:uppert}), and (\ref{eq:zfl}), to confirm the bound in Eq. (\ref{eq:zfllim}).  

Let us now study the behavior of the $h_c(f)$ in the high frequency regime, $f \gg f_c$. A good starting point is the derivative  $\dot{h}$, Eq. (\ref{dothexpr}). 
On physical grounds we know that the luminosity $L_{\nu}(t)$ is positive and bounded from above. 
Furthermore, it is natural to assume that product $L_\nu(t) \alpha(t)$
is zero outside a finite interval of time. 
This because the \n\ burst has a characteristic duration of tens of seconds (with a sharp decline of $L_\nu(t)$ at $t\sim 30-40$ s post-bounce, when the \n\ emission transitions from surface to volume emission). Furthermore, numerical simulations (see for example \cite{ Kotake_2007,Kotake:2009rr,Vartanyan:2020nmt}) suggest that the anisotropy parameter $\alpha(t)$ be mainly due to the spiral SASI, which has a characteristic duration of ${\mathcal O}(10^{-1})$ s. 

These arguments justify us to treat  $\dot{h}(t)$ as a function that has compact support in a given time interval:
\be
\begin{cases} 
      \dot{h}(t)\neq 0 & t_1\leq t \leq t_2 \\
      \dot{h}(t)= 0 & \text{otherwise} 
\end{cases}~. 
\ee

We can then use one of the incarnations of the Paley–Wiener theorem, which asserts (see for example \cite{10.2307/j.ctt1bpm9w6}): 

\vspace{0.5cm}
{\it The Paley–Wiener theorem}: {\it Let $g(t)$ be a $C^{\infty}$ function vanishing outside an interval $[A,B]$, then $\Tilde{g}(f)$ is an analytic function of exponential type $\sigma = {\rm max}\{|A|,|B|\}$\footnote{An analytic function is said to be of exponential type $\sigma$ if for every $\epsilon>0$ there exists a real constant $A$ such that $|\tilde{g}(z)|\leq A\, e^{(\sigma+\epsilon)|z|}$ for $|z|\to\infty$} and is rapidly decreasing, i.e, $|\Tilde{g}(f)|\leq c_N \left(1+\frac{f}{f_c}\right)^{-N}$ for all $N$, where $f$ is the frequency in the present context.}
\vspace{0.5cm}
Which immediately implies that \footnote{Notice that the constant $c_N$ has to be positive and $f_0$ is the typical frequency scale of the particular model. },
\be \label{PWT}
|\Tilde{\dot{h}}(f)|\leq c_N \left(1+\frac{f}{f_c}\right)^{-N}
\ee
Using \eqref{eq:dotfourier}, we have
\be
h_c(f)=\frac{|\Tilde{\dot{h}}(f)|} {\pi}\leq \frac{c_N}{\pi} \left(1+\frac{f}{f_c}\right)^{-N}\,,
\ee
for all integers $N$. By staying as conservative as possible, we can take the less constraining integer $N=1$. In principle, as the theorem states, $c_N$ can be any constant that allows the bound to exist, but we can estimate it in our case, by comparing it with the zero-frequency limit, in other words, taking $f=0$ in the expression above, we have,
\be
h_c(f)\leq \frac{c_1}{\pi} \,,
\ee
which combined with \eqref{eq:zfl} allow us to state that the high-frequency behaviour of the characteristic strain should satisfy the decaying property, 
\be
h_c(f)\leq \frac{|\Delta h|}{\pi} \left(1+\frac{f}{f_c}\right)^{-1}\,.
\ee
This result provides us with a nice interpolation between the zero frequency limit -- which leads us to the flat bound (frequency-independent) in Eq. \eqref{eq:zfllim} -- and a increasingly stringent  bound at higher frequencies. Such trend will be confirmed in all our phenomenological models, as will be seen in Sec. \ref{sec:results}.

\section{A phenomenological model of neutrino memory}
\label{sec:pheno}

In this section we construct a phenomenological model for the \mm\ effect, first by taking inspiration from the results of numerical simulations, and then generalizing to a broader range of situations. 
To keep the model sufficiently simple, in its analytical form, we will concentrate on reproducing the features of the \n\ luminosity and of the anisotropy parameter that develop over time scales of $0.1$ s or larger. These correspond to frequency scales ($f \lesssim 10$ Hz) at or close to the Deci-Hz range, which is most promising experimentally.

\subsection{Neutrino luminosity and \ans\ parameter}
\label{subsec:general}

This subsection contains a brief overview of \n\ emission from a \sn\ -- with emphasis on the aspects most relevant to the \mm\ --  for the benefit of the broader readership. \\

The aging process of a massive star ($M \geq 8 \msun$) involves several phases of nuclear burning,  finally culminating in a pressure loss, which  leads to the gravitational collapse of the star's core. Due to a sharp rise in the incompressibility of nuclear matter, the collapse eventually  comes to an abrupt stop, and the core bounces back, producing a forward moving shockwave. The shockwave is initially stalled for a fraction of a second, and then it either dies out (leading to black hole formation) or is launched due to energy deposition by \ns, thus causing the explosion of the star.  

As a result of the core collapse and bounce, $E_{tot} \sim  3~10^{53}$ ergs of gravitational energy is released, and most of it is emitted in thermal neutrinos and anti-neutrinos of all flavors, over a time scale of $\sim 1-10$ s. The emission is largely isotropic, however, transient anisotropies of the order of $\sim 10^{-3}-10^{-2}$ are expected to develop. 
We can distinguish three main phases for the \n\ emission: 

\begin{itemize}

   \item The \emph{neutronization burst}.  
The initial emission of \ns\ after core collapse is dominated by electron neutrinos ($\nue$) due to the absorption of electrons on neutrons and nuclei. 
The signature of this processes is a sharp peak in $L_\nu(t)$, of about $\sim 2$ ms width. 
Numerical simulations \cite{Kotake_2007,Kotake:2009rr,Vartanyan:2020nmt} show that, at this stage, the anisotropy parameter is negligible ($\alpha (t) \leq 0.001$); an indication that the shock maintains spherical symmetry.  Here we will assume $\alpha=0$ during neutronization. 

\item The \emph{accretion phase}. 
Until $t\sim 0.2 - 1.0$ s post-bounce, when the shockwave is stalled, the \n\ emission is approximately thermal and it is powered by the influx of matter accreting on the collapsed core. The \n\ luminosity time profile, after the sharp neutronization peak, becomes nearly flat, reaching a plateau value  of $L_\nu \simeq few~10^{52}~{\rm erg~ s^{-1}}$. 
Numerical simulations confirm that in the accretion phase the physics near the core is complex, being characterized by turbulence and hydrodynamical instabilities, like the Standing Accretion Shock Instability (SASI), which causes fluctuations of the \n\ luminosity around the plateau value with a characteristic time scale $\delta t \simeq 10^{-2}$ s \cite{Blondin:2006fx,Kotake_2007, Kotake:2009rr,Walk_2020}.
The same phenomena lead to anisotropies in the \n\ emission; 
in particular, the \emph{spiral} SASI mode has been found to be associated to an anisotropy parameter  $|\alpha| \sim 10^{-3} - 10^{-2}$ \cite{janka_muller,Kotake:2009rr}.  $\alpha$ could change sign over time, transitioning between positive and negative, as the orientation of plane of the spiral SASI changes relative to the observer \cite{Kotake:2009rr}; see Fig.~\ref{fig:nac2} for an example.   

\item The post-accretion time: \emph{cooling phase}. If the shock is launched, the \n\ emission continues beyond the accretion phase (otherwise, in the case of black hole formation, it drops sharply at $t\sim 0.5 - 1 $ s post-bounce, see e.g., \cite{Kuroda:2017trn,Kuroda:2018gqq,Vartanyan:2019ssu,Walk_2020,Shibagaki:2020ksk}). The collapsed core -- which is now a newly-born proto-neutron star -- and its surrounding regions  slowly cool by thermally radiating neutrinos of all flavors. The \n\  luminosity and average energy decline smoothly with time, over a scale of  ${\mathcal O}(10)$ s. Since the state-of-the-art numerical simulations stop at or before the end of the accretion phase, there are no quantitative estimates of the anisotropy parameter in the cooling phase. Intuitively, one may expect a relaxation of the system into a more symmetric configuration (smaller anisotropy), however the question remains open. 

\end{itemize}

\subsection{A model: phenomenological description of the \n\ memory}
\label{subsec:template}

Let us now construct a phenomenological description of the \mm\ strain that well approximates the results of numerical simulations, and can serve as a template for generalizations to a wider range of cases (Sec. \ref{sec:results}).  

\begin{figure}
    \begin{center}
        \subfloat[]{\includegraphics[width=0.65\textwidth]{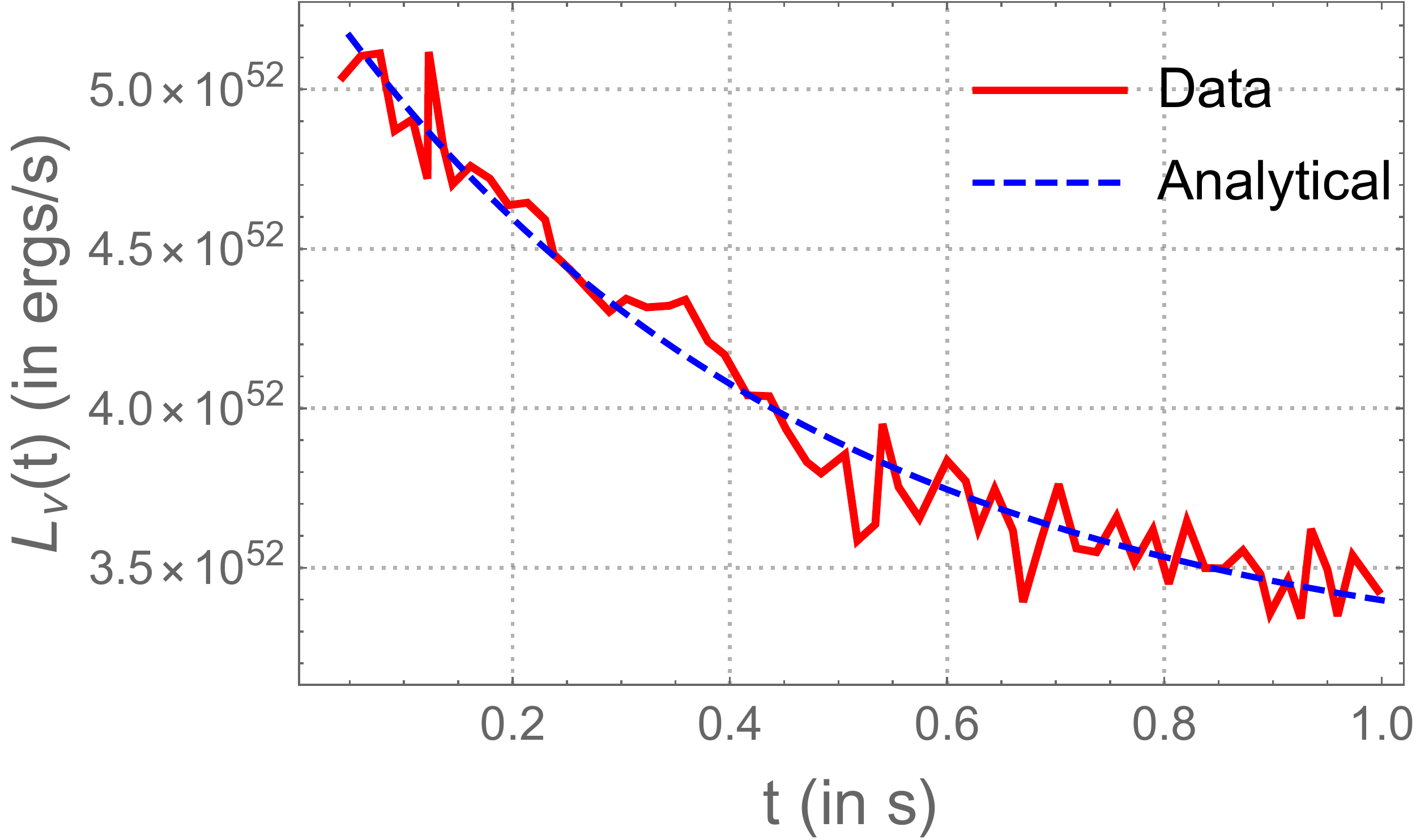}\label{fig:nac1}} \\
        \subfloat[]{\includegraphics[width=0.59\textwidth]{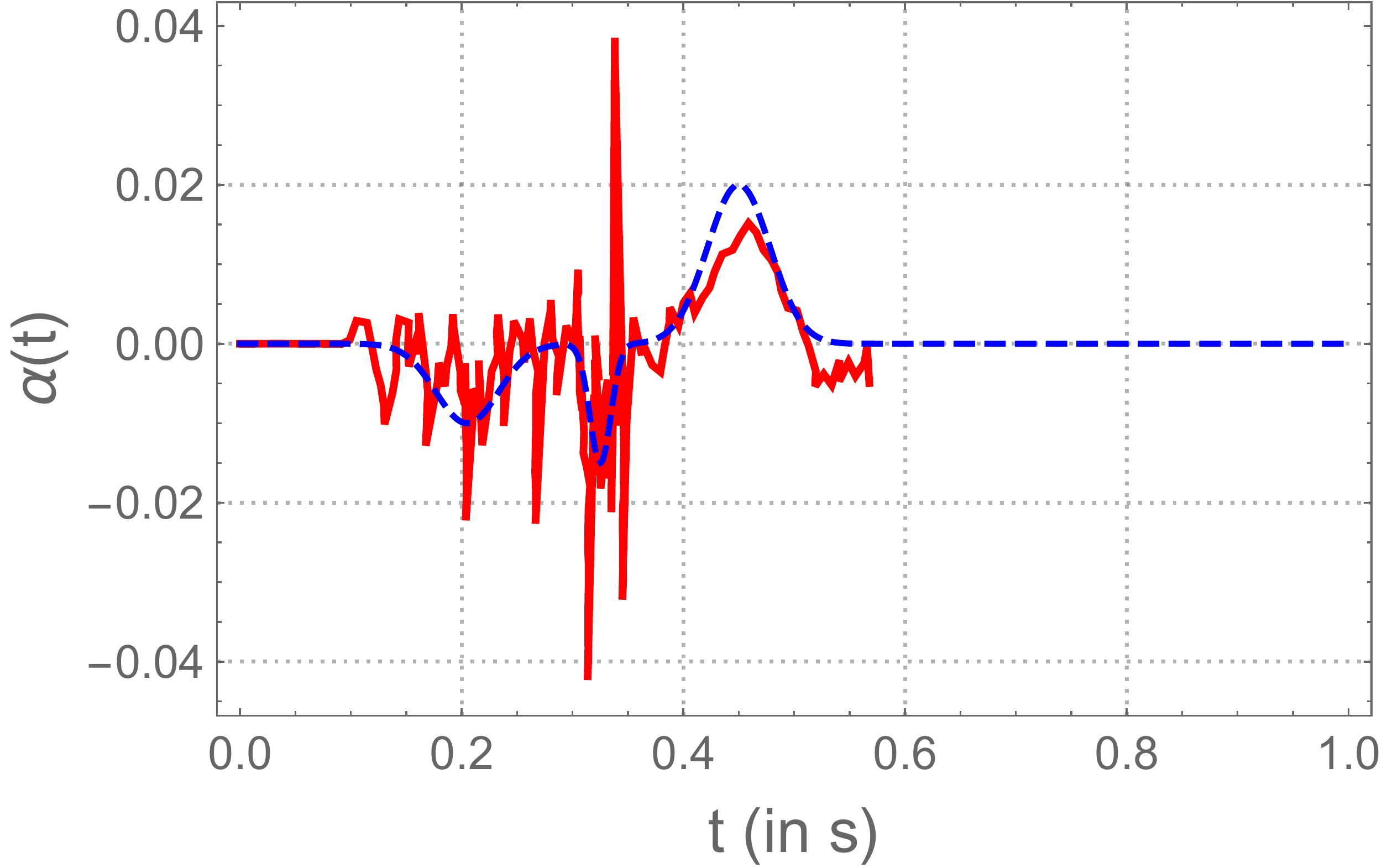}\label{fig:nac2}} \\            \subfloat[]{\includegraphics[width=0.65\textwidth]{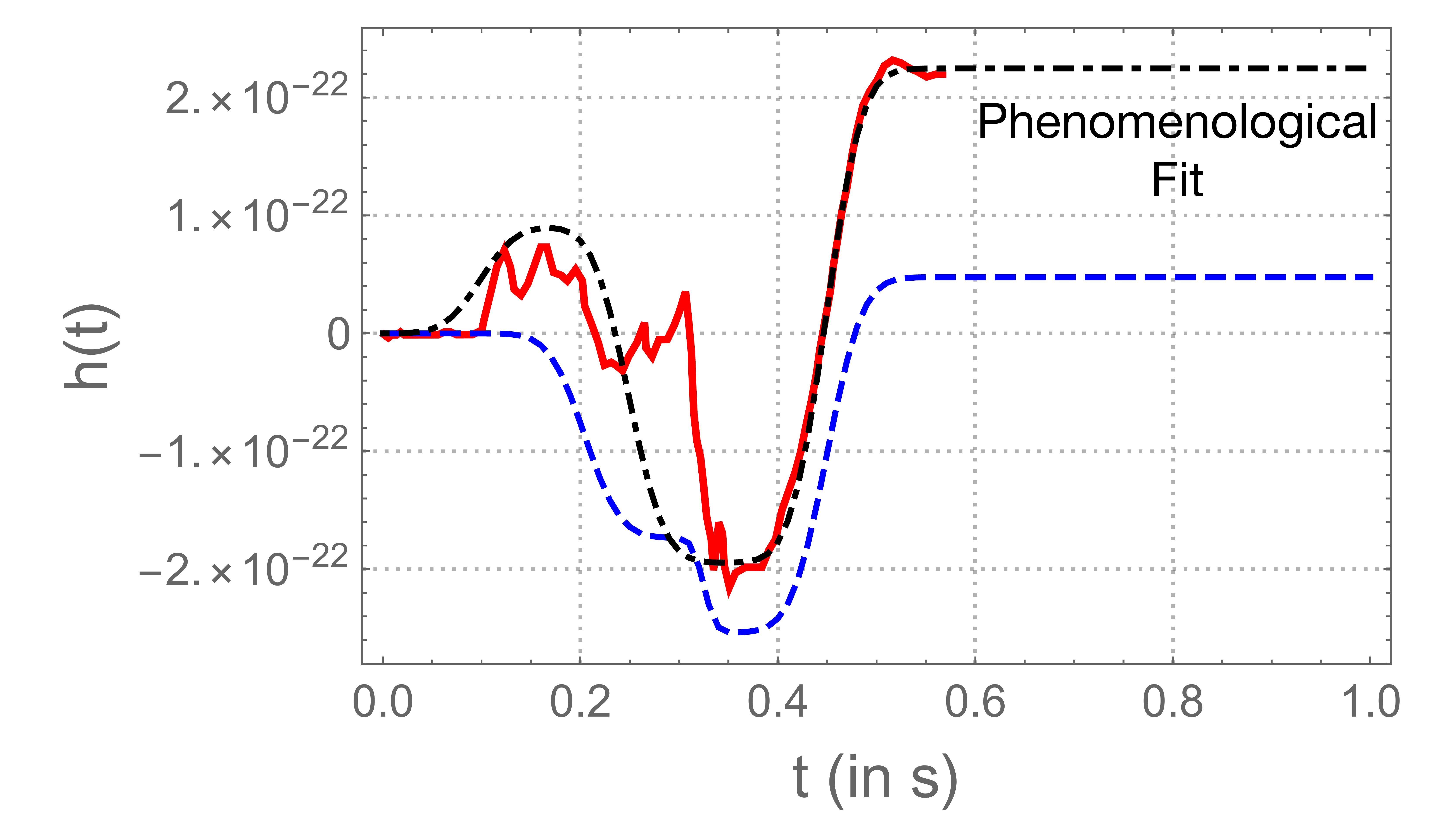}\label{fig:nac3}}
    \end{center}
    \caption{\label{fig:num_ana_compare}  Plots showing the phenomenological curves from our model (dashed lines), in comparison with numerical results (solid lines) for: (a) the \n\ ($\nu_e$) luminosity, $L_\nu (t)$; (b) anisotropy parameter $\alpha (t)$; and (c) the \gw\ strain, $h (t)$. The numerical results in (a) are from ~\cite{Vartanyan:2020nmt}, whereas those in (b) and (c) are from \cite{Kotake:2009rr} (in all cases graphics is adapted from the original papers, with the permissions of the authors). 
    In (c), the dashed (blue) curve is obtained using the same parameters as the curves in (a) and (b). An additional line (dot-dashed, black) is plotted, showing a phenomenological fit to the numerical data using Eq.~\eqref{eq:mastereq}. 
    }
\end{figure}

To fix the ideas, we first consider a scenario where the \n\ emission has \ans\ only in the accretion phase, and therefore only this phase contributes to the \mm\ effect. As discussed in Sec. \ref{subsec:general}, 
during the accretion phase $L_\nu(t)$ has an irregular behavior over time scales $\delta t \sim 0.01 $ s or so, due to turbulence and SASI, however its global shape over a time interval $\Delta t \gtrsim 1$ s  -- which is most relevant to capture the spectrum at low frequency, as will be shown later -- can be approximately described by the  functional form  
\be
L_{\nu} (t) = \lambda + \beta\ \exp{\big(- \chi\,t \big)}\,~,
\label{eq:lumform}
\ee
(see Fig.~\ref{fig:num_ana_compare}). Here $l$, $\beta$ and $\chi$ are phenomenological parameters, and it is assumed that Eq. (\ref{eq:lumform}) is only valid locally (i.e., for a finite interval of time post-bounce), since $L_{\nu} (t)$ should vanish at  $t\rightarrow \pm \infty$.  

Inspired by published numerical results \cite{Kotake_2007,Kotake:2009rr,Vartanyan:2020nmt,Suwa_2009}, we model the asymmetry function as a multi-Gaussian, added to a constant component: 
\be
\alpha (t) = 
   \kappa + \sum_{j=1}^N \xi_j\ \exp{\Bigg( -\frac{(t-\gamma_j)^2}{2\sigma_j^2} \Bigg)}~, 
\label{eq:alphaform}
\ee
for $t>0$ \footnote{For the sake of obtaining closed analytical formulae for the \mm\ in time and frequency space, we imposed $\alpha(t)=0$ for $t\leq 0$; this is immaterial for the conclusions of this work.}.
While largely oversimplified, this form captures the essential physical features of $\alpha(t)$ (Sec. \ref{subsec:general}), as is shown in Fig.~\ref{fig:num_ana_compare}, where Eq. (\ref{eq:alphaform}) is compared to a numerical result. 
Furthermore, Eqs. (\ref{eq:lumform}) and (\ref{eq:alphaform}) lead to reasonably accurate results for the \mm\ strain, as is discussed below. 

By substituting the expressions  \eqref{eq:lumform} and \eqref{eq:alphaform}  in Eq. \eqref{eqn:mainampeq}, one obtains a closed form for the \mm\ strain as a function of time: 
\begin{align}
h(t) &= \sum_{j=1}^N \Bigg\{  
\left[ h_{1j} \Bigg( \erf{( \rho_{j}\ \tau_{1j} )}+\erf{\Big( \rho_{j}(t-\tau_{1j})\Big)} \Bigg) 
\right]
+ \left[ h_{2j}\Bigg( \erf{( \rho_{j}\ \tau_{2j} )} + \erf{\Big(\rho_{j} (t-\tau_{2j}) \Big)} \Bigg) \right]\Bigg\} \nonumber \\ 
&+ \left[ h_{3} \Bigg( \frac{\beta}{\chi} \ \Big( 1-\exp{(-t\chi)} \Big) + \lambda t \Bigg) \right]
\,,
\label{eq:mastereq}
\end{align}
where, 
\begin{align}
\label{eq:htformb}
h_{1j}&= \frac{2G}{rc^4} \sqrt{\frac{\pi}{2}} \beta \xi_j \sigma_j  \exp\Big( \frac{\chi}{2} (-2\gamma_j + \sigma_j^2 \chi) \Big)~, \nonumber\\
\rho_{j}&=\frac{1}{\sqrt{2} \sigma_j}\,,\nonumber\\
\tau_{1j}&= \gamma_j - \sigma_j^2 \chi \,,\\
h_{2j}&= \frac{2G}{rc^4} \sqrt{\frac{\pi}{2}}  \lambda \xi_j \sigma_j \,,\nonumber\\
\tau_{2j}&=  \gamma_j\,,\nonumber\\
h_{3}&= \frac{2G}{rc^4} \kappa\,.\nonumber
\end{align}
Here $\erf{(x)}$ is the error function, $\erf{(x)} = \frac{2}{\sqrt{\pi}} \int_0^x \exp{(-t^2)} dt$, and a factor of ${2G}/{rc^4} = 5.34 \times 10^{-65}~(10 {\rm ~kpc}/r)$ m has been isolated where possible, to facilitate the comparison with Eq.  \eqref{eqn:mainampeq}.  

In the expression \eqref{eq:mastereq}, we can identify the main features of the \mm\ signal. For a single Gaussian anisotropy bump ($N=1$), there are three distinct terms. The first and second terms are due to the exponential and constant part of $L_\nu$, respectively, in combination with the Gaussian part of $\alpha(t)$. They show the typical rise and plateau behavior expected from the theory (Sec. \ref{subsec:genprop}, Fig. \ref{fig:ameba}), where the timescale of the rise is, naturally, given by the  width of the anisotropy Gaussian, $\sigma_j$.  The third term is proportional to the constant part of the anisotropy parameter, and therefore the time-scale of its rising and plateauing is the time-scale of the \n\ luminosity, $1/\chi$. 
Finally, in Eq. \eqref{eq:mastereq} one may notice a term of the form $\kappa \lambda t$, which is due to the constant terms in $L_\nu$ and in $\alpha(t)$; this term vanishes in realistic realizations (see next section), and therefore  it is not a cause of concern.

Let us now describe the \mm\ strain in the frequency domain. 
From Eq. (\ref{eq:mastereq}), a closed form is obtained for the Fourier transform of $h$:
\begin{align}
\tilde{h} (f) &= \sum_{j=1}^N \Bigg[ \Bigg(  \ h_{1j} \frac{i}{\pi f} \exp\Big( \frac{- \pi^2 f^2}{\rho_{j}^2} \Big) \exp \Big( i 2 \pi f \tau_{1j}  \Big) \Bigg)  + \Bigg( h_{2j}  \frac{i}{\pi f} \exp\Big( \frac{- \pi^2 f^2}{\rho_{j}^2} \Big) \exp \Big( i 2 \pi f \tau_{2j}  \Big) \Bigg)\Bigg] \nonumber \\ &+ \Bigg(  \sqrt{2\pi}\ h_{3} \frac{\beta}{\chi}\ \Big( \frac{1}{i 2 \pi f} - \frac{1}{-\chi + i 2\pi f} \Big)  \Bigg) \,,
\label{eq:mastereq2}
\end{align}
where $i$ is the imaginary unit. 

Here we analyze the structure of Eq. (\ref{eq:mastereq2})  to infer the  properties of $h_c(f)=f |\tilde{h} (f)|$. In Eq. \eqref{eq:mastereq2}, the terms proportional to $f^{-1}$ produce the 
expected low-frequency limit, where $h_c(f)$ tends to a constant value (see Sec. \ref{subsec:ZFLHFL}).  We also observe that $f \tilde{h} $ vanishes in the high frequency limit, thus reproducing the  expected drop, $h_c (t)\rightarrow 0$. 
The the transition between the two regimes (low and high $f$ limits) is determined by the inverse width of the Gaussian asymmetry factors, $f_i \approx 1/2 \pi \sigma_j$ (first two terms of Eq. \eqref{eq:mastereq2}) or, in the case of constant asymmetry, by the inverse time scale of the \n\ luminosity, $f_\nu \approx \chi/2\pi$. 

Let us now give an illustration of how our model reproduces the expected features of a \mm\ signal. In Fig.~\ref{fig:num_ana_compare}  we compare the phenomenological forms for $L_\nu$, $\alpha(t)$ and  $h(t)$ (Eqs. (\ref{eq:lumform}), (\ref{eq:alphaform}) and (\ref{eq:mastereq}), respectively) with the results of numerical simulations for the accretion phase. Due to the sparseness of published numerical results, we consider information from different sources, and in particular, $L_\nu(t)$ from \cite{Vartanyan:2020nmt} (see Fig. 1 (left) there, $15\msun$ model) and $\alpha(t)$ from \cite{Kotake:2009rr}. These are well reproduced, in their global structure, by the phenomenological curves (with a tri-Gaussian structure for $\alpha(t)$, $N=3$), for appropriately chosen parameters (given in Table~\ref{tab: fit_parameters}). 
For the \emph{same} parameters, the \mm\ strain, $h(t)$ from Eq. \eqref{eq:mastereq} is plotted in Fig.~\ref{fig:nac3}. For comparison, the figure also shows the numerically calculated $h(t)$ from ~\cite{Kotake:2009rr}.
The two curves are in good qualitative agreement in the general time structure of the strain, although differences by a factor of up to $\sim 4-5$ exist. 
We stress that a quantitative agreement is not expected because of our extracting information on $L_\nu$ and $\alpha$ from different sources, therefore the qualitative agreement noted above is a satisfactory validation of our model.

Our master formula, Eq. \eqref{eq:mastereq}, can serve an effective phenomenological description of data (or numerically-generated results) for $h(t)$, if its parameters are treated as fit parameters. 
The dot-dashed curve in Fig.~\ref{fig:nac3} shows an example of this: the dot-dashed curve has been obtained from Eq. \eqref{eq:mastereq} (with $N=3$) by setting the parameters so to best reproduce the numerically calculated $h(t)$ from Ref. \cite{Kotake:2009rr}.  The agreement is acceptable, and can be further improved by increasing $N$. 
\\

As a note in closing, let us mention that our phenomenological forms for $h(t)$ and $\tilde h(f)$ are in general agreement with earlier, simpler toy models, like those presented by M.~Favata \cite{Favata:2011qi}. In those, $h(t)$ has the form of a hyperbolic tangent. A comparison with our Eq. \eqref{eq:mastereq} becomes intuitive if one considers the well known approximation (see, e.g. \cite{doi:10.1119/1.15018}):
\begin{equation}
    \erf{(x)} \simeq \tanh{(mx)}, \text{ with }m = \sqrt{\pi} \log (2)\,,
    \label{eq:tanhapprox}
\end{equation}
which we checked to be very precise (less than 1\% difference) at the time/frequency regimes of interest here.

Using Eq. (\ref{eq:tanhapprox}), our master equation for $h(t)$, Eq.~\eqref{eq:mastereq}, can be rewritten as:
\begin{align}
h(t) &= \sum_{j=1}^N \Bigg[ \Bigg\{ h_{1j} \Bigg( \tanh{( m\rho_{j}\ \tau_{1j} )}+\tanh{\Big( m\rho_{j}(t-\tau_{1j})\Big)} \Bigg) \Bigg\} + \Bigg\{ h_{2j}\Bigg( \tanh{( m\rho_{j}\ \tau_{2j} )} \nonumber \\ &+ \tanh{\Big(m\rho_{j} (t-\tau_{2j}) \Big)} \Bigg) \Bigg\}\Bigg] + \Bigg\{ h_{3} \Bigg( \frac{\beta}{\chi} \ \Big( 1-\exp{(-t\chi)} \Big) + \lambda t \Bigg) \Bigg\} \,,
\label{eq:mastereqtanh}
\end{align}
and its Fourier transform takes the form:
\begin{align}
\tilde{h} (f) &= \sum_{j=1}^N \Bigg[ \Bigg( \ h_{1j} \frac{i \pi}{m\rho_j} \csch\Big( \frac{\pi^2 f}{m\rho_j} \Big) \exp \Big( i 2 \pi f \tau_{1j}  \Big) \Bigg)  + \Bigg( h_{2j}  \frac{i \pi}{m\rho_j} \csch\Big( \frac{\pi^2 f}{m\rho_j} \Big) \exp \Big( i 2 \pi f \tau_{2j}  \Big) \Bigg) \Bigg] \nonumber \\ &+ \Bigg(  \sqrt{2\pi}\ h_{3} \frac{\beta}{\chi}\ \Big( \frac{1}{i 2 \pi f} - \frac{1}{-\chi + i 2\pi f} \Big)  \Bigg)\,. 
\label{eq:mastereqtanh2}
\end{align}

\section{Generalization: plausible phenomenological scenarios}
\label{sec:results}

We now present five models that are a generalization of the framework discussed in Sec.~\ref{subsec:template}.  These models represent possibilities that have not yet been investigated numerically, but are nevertheless plausible. In the remainder of this section, for each model we provide some physics motivation, and justify the choice of the parameters (given in Table~\ref{tab: parameters}).

\subsection{Case studies}
\label{subsec:models}

\begin{figure}
    \begin{center}
        \subfloat[]{\includegraphics[width=0.6\textwidth]{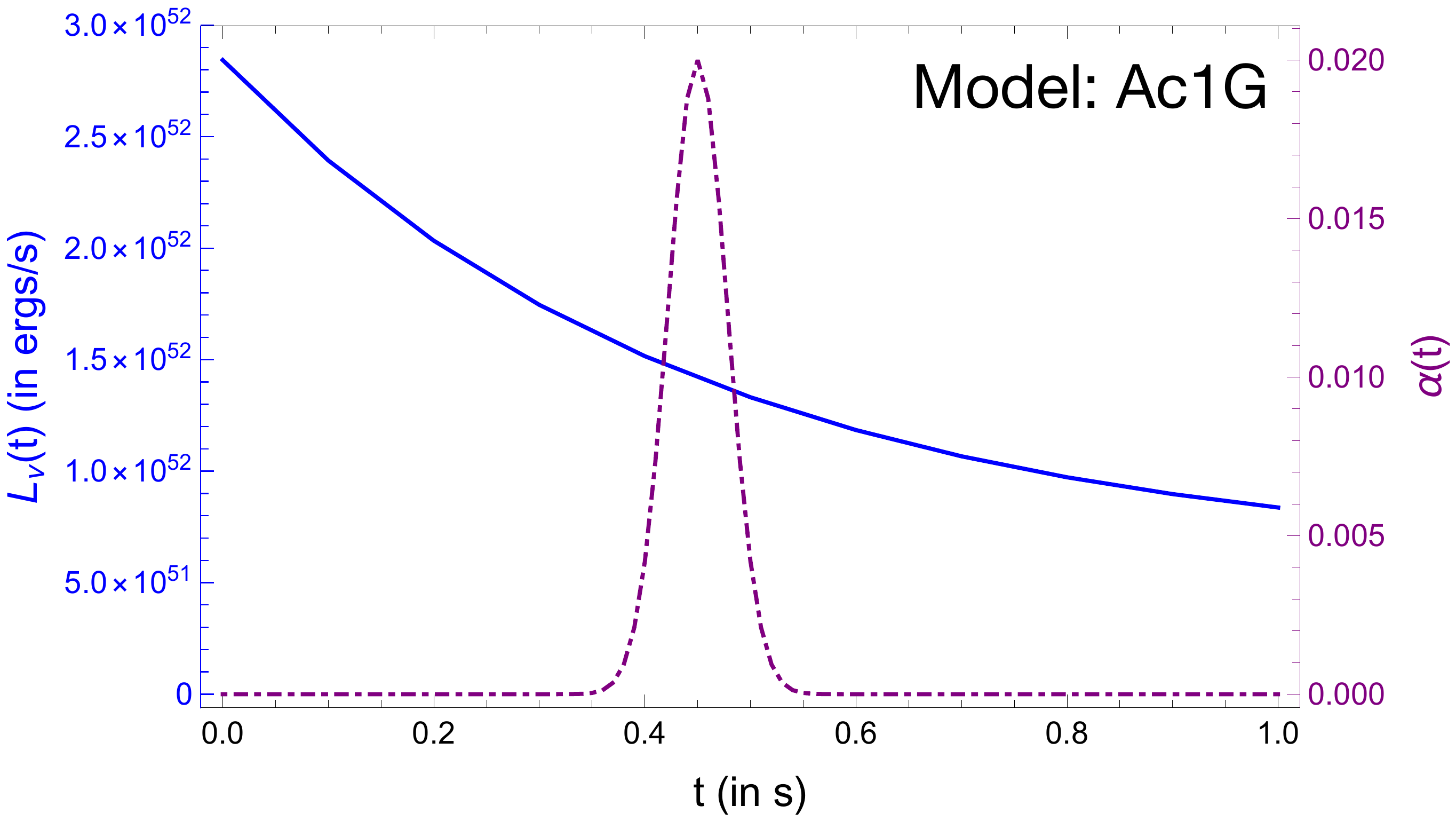}\label{fig:mod2}} \\        
        \subfloat[]{\includegraphics[width=0.6\textwidth]{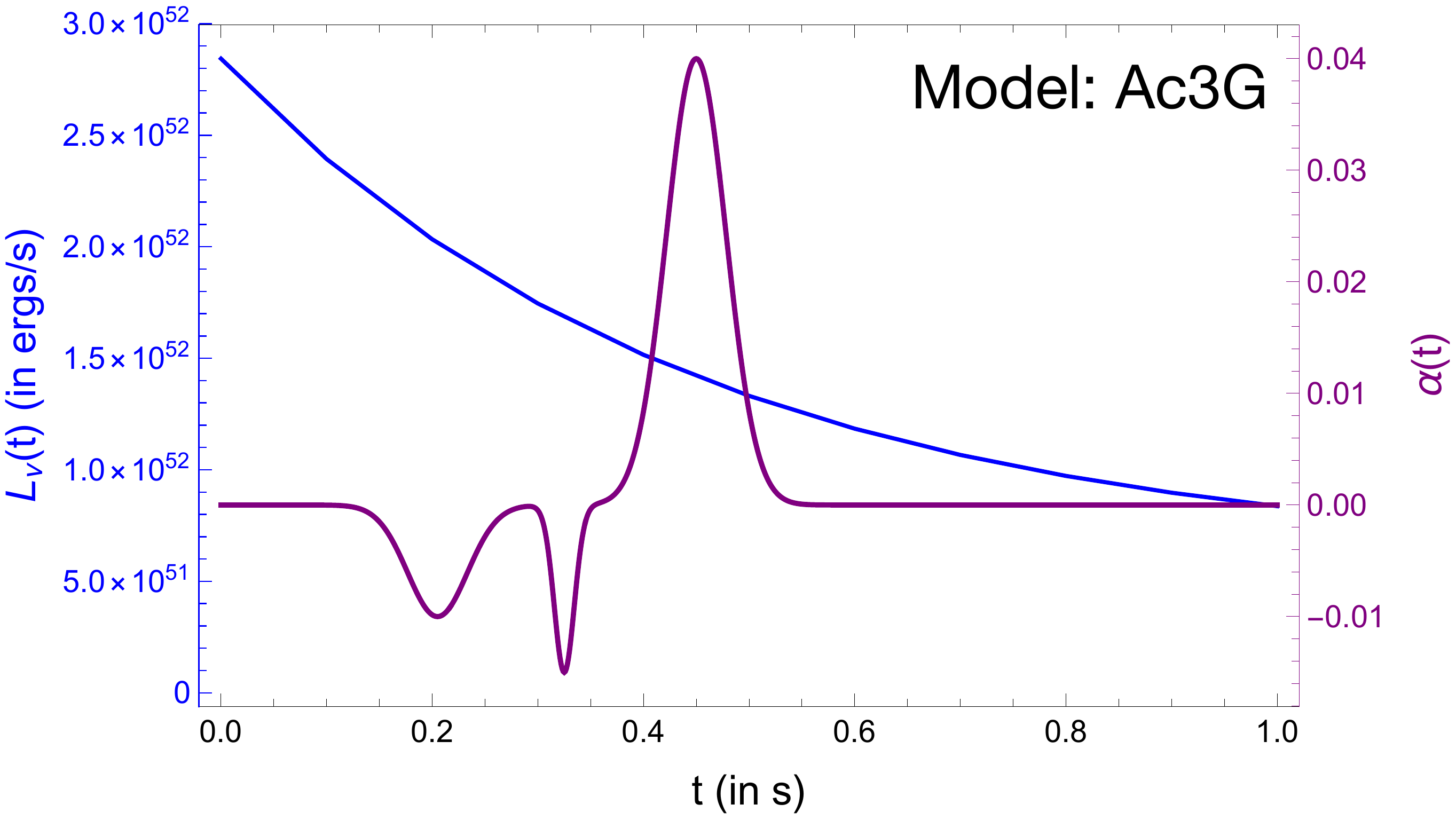}\label{fig:mod3}}\\
        \subfloat[]{\includegraphics[width=0.6\textwidth]{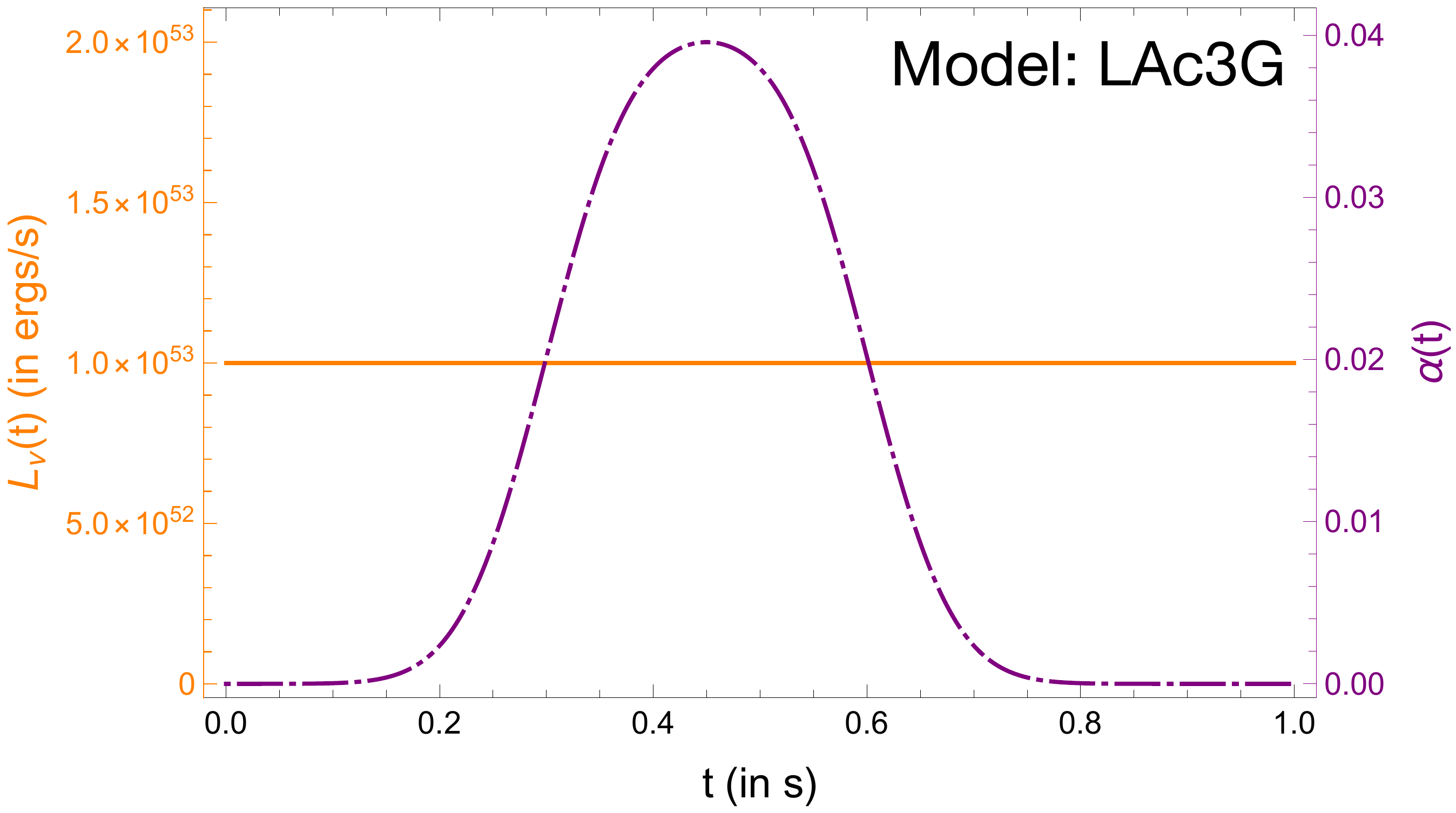}\label{fig:mod5}}
    \end{center}
    \caption{\label{fig:models2_3} Neutrino luminosity (exponential curve, vertical scale on the right) and anisotropy parameter (single- or multi-Gaussian curve, vertical scale on the left) as functions of time post-bounce, for the three accretion-only models, where only the accretion phase of the \n\ emission contributes to the memory.  See Table \ref{tab: parameters} for details. 
    }
\end{figure}

To fix the ideas, let us examine the following specific situations: 

\begin{itemize}
    \item \emph{Accretion-only models.} Taking direct inspiration from numerical results (Sec. \ref{sec:pheno}), we present three realizations where only the accretion phase contributes to the \n\ \mm. They are illustrated in Fig. \ref{fig:models2_3}. The first model, the \emph{accretion phase-three gaussians model} (Ac3G), has already been introduced in Sec. \ref{sec:pheno} and Fig.~\ref{fig:num_ana_compare} as a description of numerical simulations, and therefore can be considered especially well motivated. 
 
    A variation of the previous model is the \emph{accretion phase-one Gaussian model} (Ac1G), where the anisotropy parameter $\alpha(t)$ has a simpler time-dependence, and is described by a single Gaussian profile. This scenario may be realistic for cases where the spiral SASI activity is weaker and shorter, for example in supernovae from smaller mass progenitors, see e.g., \cite{Tamborra:2013laa,Tamborra:2014hga,Walk:2018gaw,Walk:2019ier}. 
 
    As a third realization, we consider the \emph{long accretion phase-three Gaussian model} (LAc3G), where optimistic choices of the parameters are made. 
    Here, the \n\ luminosity is kept constant at a relatively high value, and the anisotropy parameter is the sum of three overlapping Gaussian curves, extending to $t\sim 0.7$ s (refer to Table~\ref{tab: parameters} for details). The net shape of the $\alpha(t)$ function is a curve that  has a fast rising and declining time scales, similarly to the Ac1G model, but is wider than a single Gaussian. 
    This model  could be descriptive of a black-hole forming collapse (failed \sn), where the high rate of mass accretion and the long-stalling shockwave favor a highly luminous and sustained \n\ emission and prolonged spiral SASI, see for example the simulation for a  $40 \msun$ progenitor in Ref.~\cite{Walk_2020}, where a complex SASI dynamics (suggesting a multi-Gaussian structure of $\alpha(t)$) is found, and the collapse to a black hole is obtained a at $t=0.570$ s post-bounce.
    
    \begin{figure}
    \begin{center}
        \subfloat[]{\includegraphics[width=0.75\textwidth]{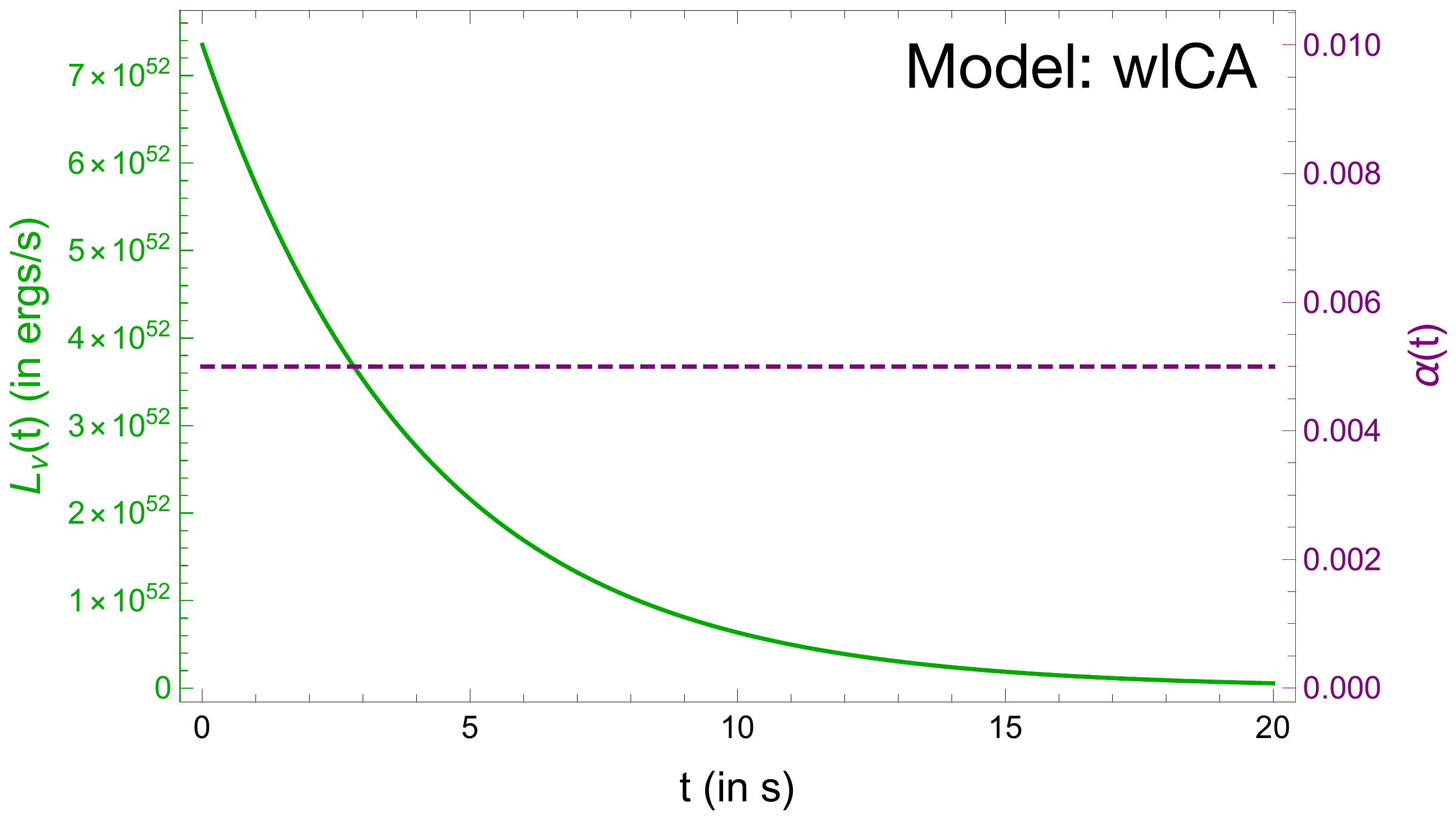}\label{fig:mod1}} 
        \\
        \subfloat[]{\includegraphics[width=0.75\textwidth]{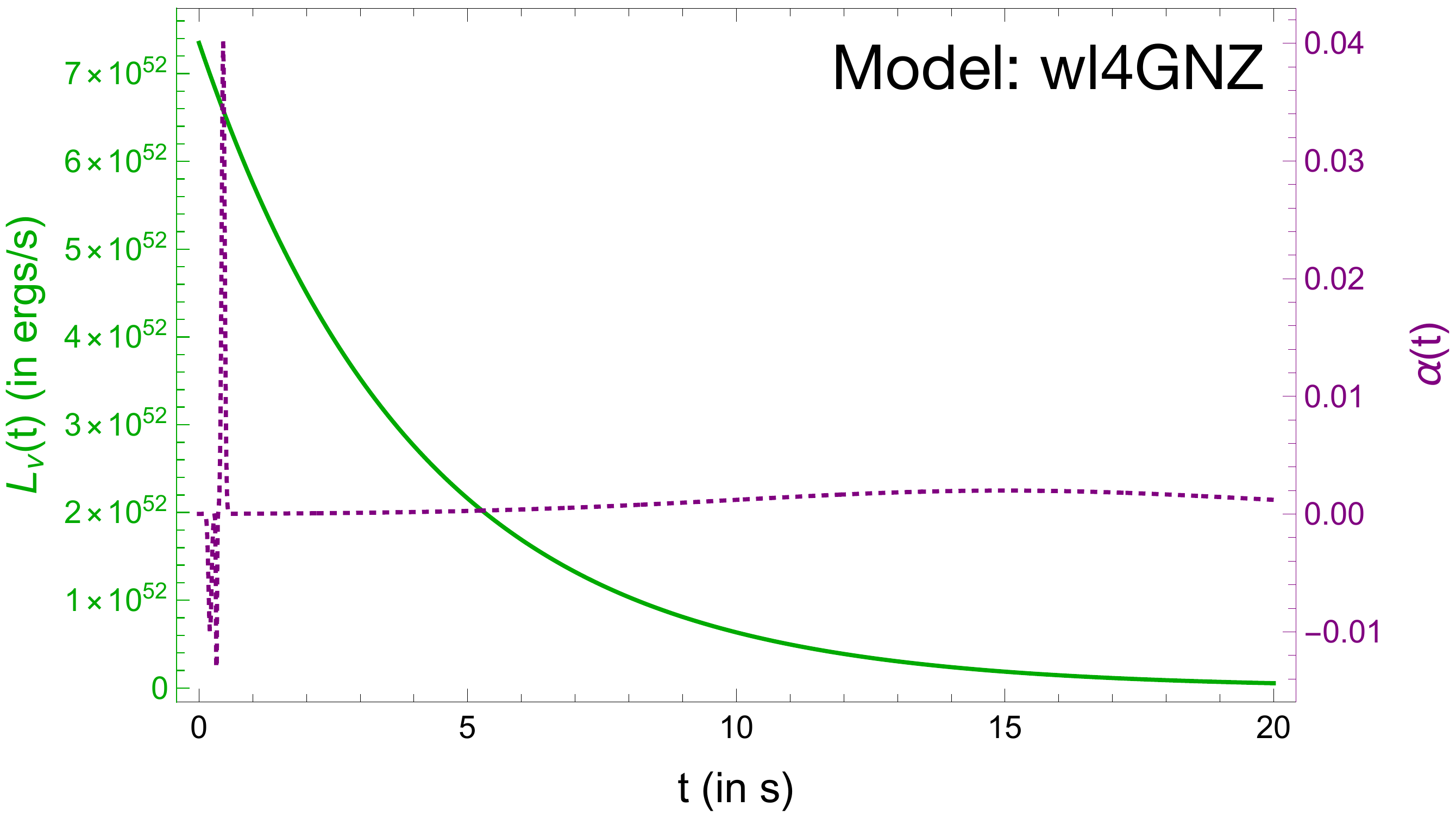}\label{fig:mod4}} \\
    \end{center}
    \caption{\label{fig:models1_4}  The same as Fig. \ref{fig:models2_3} for the whole-luminosity models, where both the accretion and cooling phases contribute to the memory. See Table \ref{tab: parameters} for details.
    }
\end{figure}

    \item \emph{Long-term evolution models.}
    We now discuss scenarios where the entire $\sim 10$ s \n\ burst contributes to the \mm\ due to a residual, non-zero long-term anisotropy, see Fig. \ref{fig:models1_4}. While the idea of a multi-second  long anisotropic emission is perhaps speculative, it is motivated by the fact that some numerical simulations  reach $t\sim 1 $ s post-bounce (where they end due to computational cost) with a non-zero $\alpha(t)$ \cite{Vartanyan:2019ssu}, thus suggesting that anisotropy could be present  beyond $t\sim 1$ s post-bounce in certain cases. 
    
    The first scenario, the \emph{whole luminosity-constant alpha model} (wlCA, show in Fig. \ref{fig:models1_4}, top pane) is the  simplest realization, having a constant anisotropy parameter.  Consistently with the idea of a residual effect, here $\alpha$ is fixed at a relatively small value ($\alpha=5~10^{-3}$) compared to the accretion-only models. A constant, feature-less anisotropy parameter could be realized if the SASI is very weak or absent and the anisotropy has a  different physical origin, i.e., in the structure of the progenitor star. This model might also be useful for comparison with prior theory works where a constant anisotropy was assumed, for example Ref. \cite{Suwa_2009}, where the case of a system composed of an accretion disk and a jet (producing a Gamma Ray Burst) was examined. 

    In the second scenario, called  the \emph{whole luminosity-four Gaussians non-zero alpha model} (wl4GNZ), we attempt a more realistic description by combining the time structure of $\alpha$ for the accretion phase ($\alpha (t)$ is the same as in model Ac3G at for $t < 1$ s), with an additional extended anisotropy, represented by a wide Gaussian centered at several seconds post-bounce, reaching a maximum value of $\alpha \simeq 0.002$. 
    Physically, this late time feature of $\alpha$ could correspond to a weak revival of the anisotropy due to late time effects. We note that here the product $L_\nu(t)\alpha(t)$ is a smoothly decreasing function of $t$ (for $t \gtrsim 1$ s). Therefore -- in the absence of a direct physical interpretation of the parameters --  this model could simply be considered as a  purely phenomenological description of a multi-second long \mm\ effect. 

\end{itemize}

\subsection{Results: memory in the time-  and frequency-domain}
\label{subsec:results}

\begin{figure}
    \begin{center}
        \subfloat[]{\includegraphics[width=0.49\textwidth]{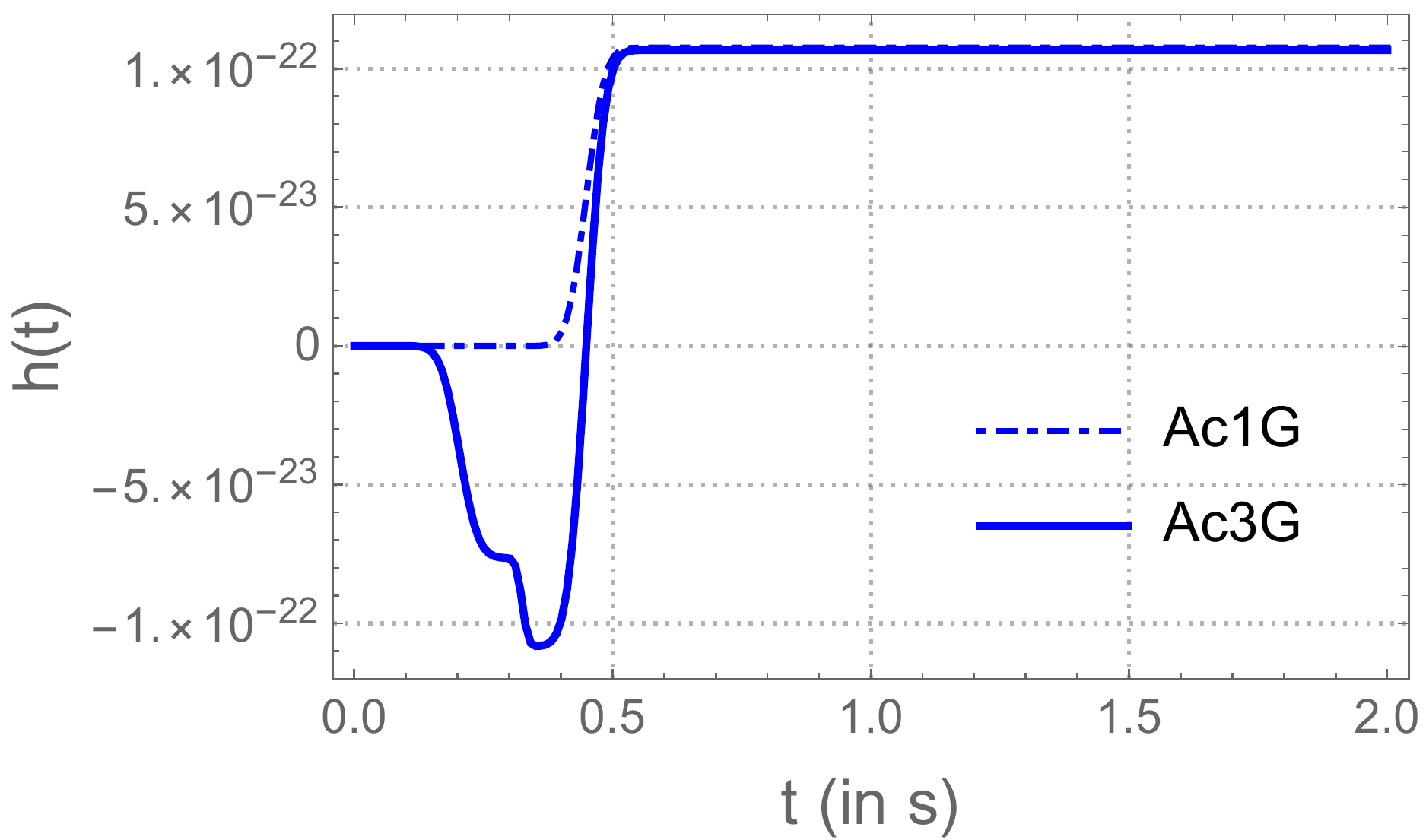}\label{fig:htp1}} 
        \subfloat[]{\includegraphics[width=0.49\textwidth]{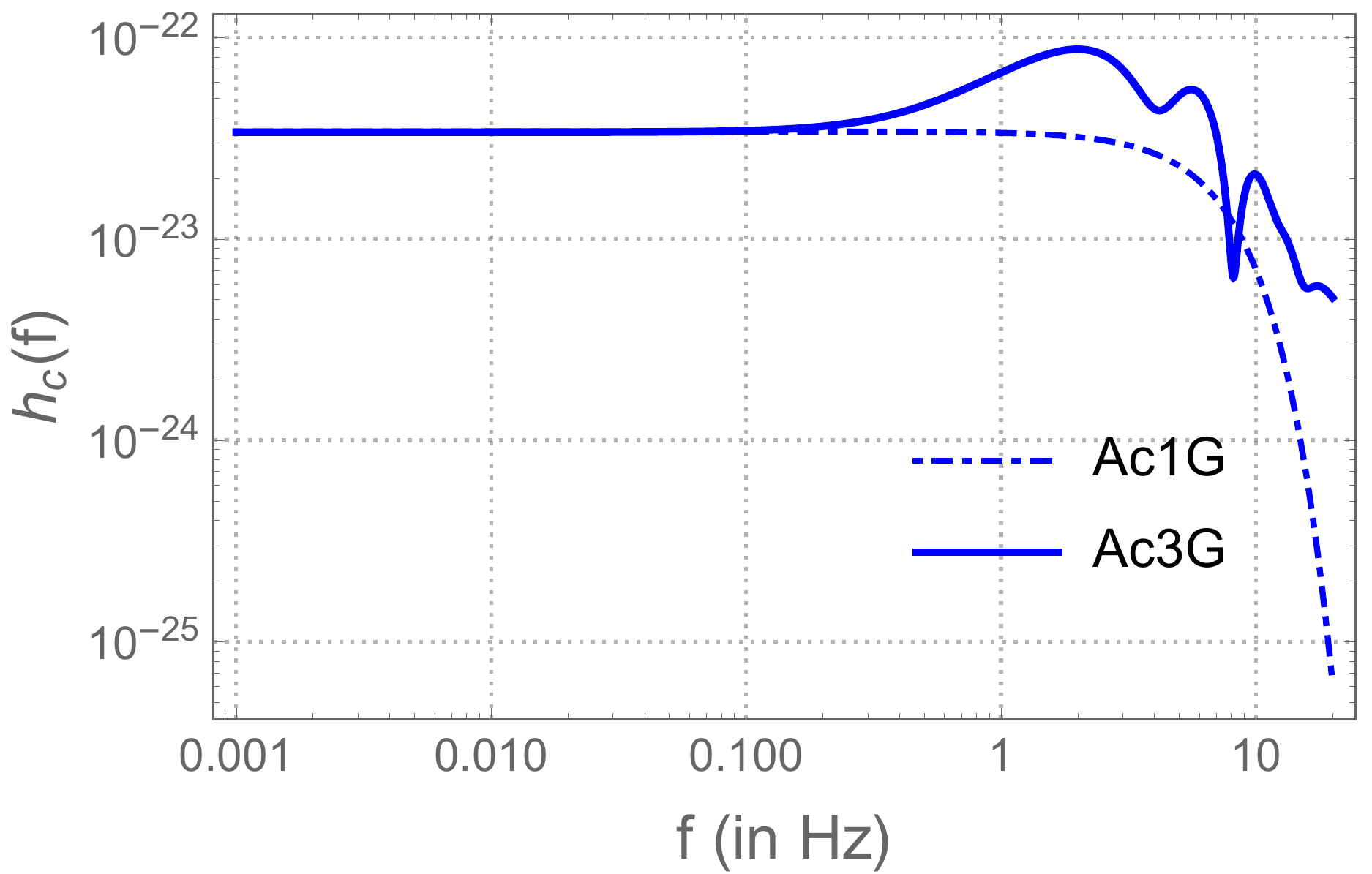}\label{fig:hcp1}}
        \\
        \subfloat[]{\includegraphics[width=0.49\textwidth]{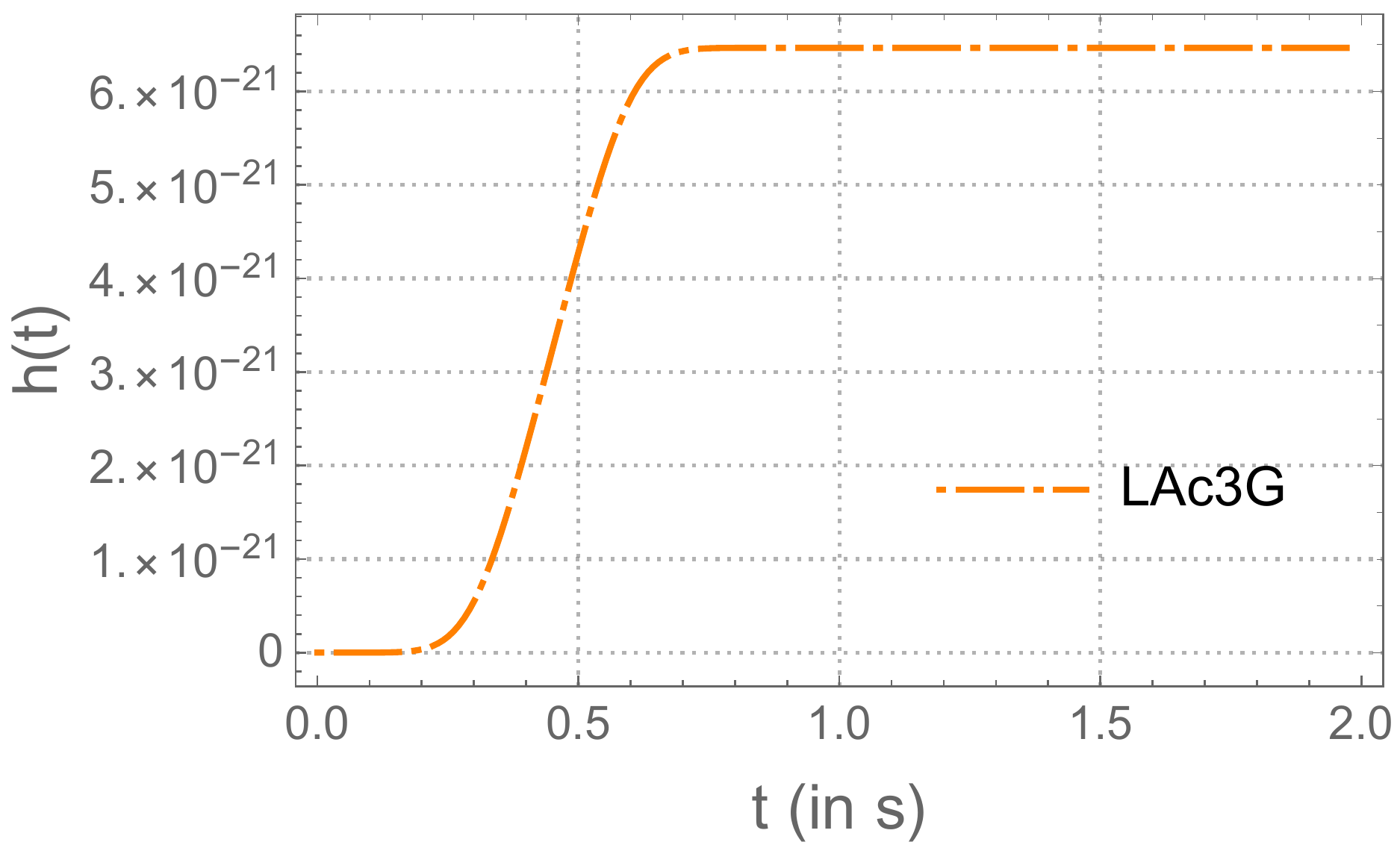}\label{fig:htp3}}
        \subfloat[]{\includegraphics[width=0.49\textwidth]{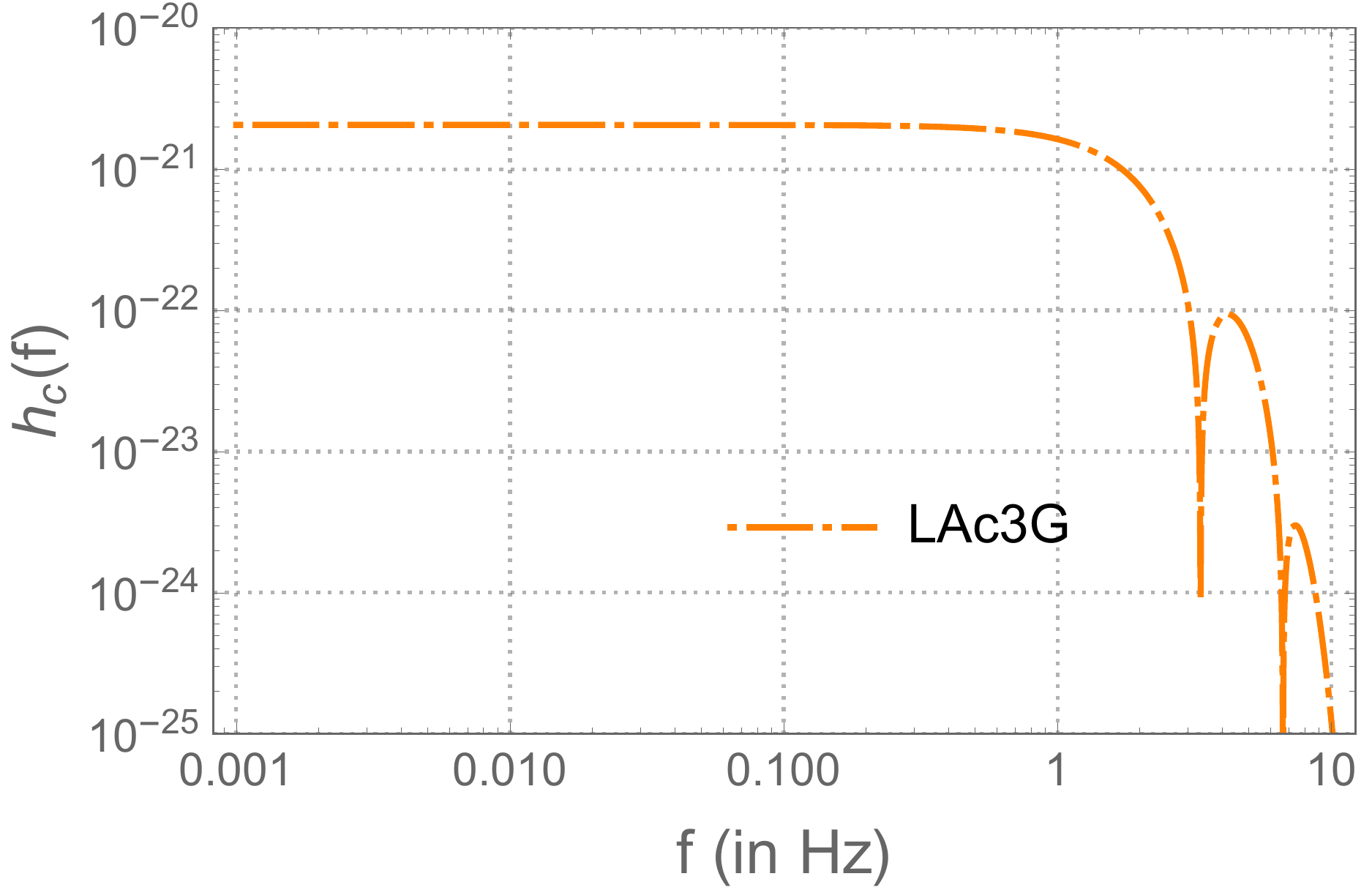}\label{fig:hcp3}}
        \\
        \subfloat[]{\includegraphics[width=0.49\textwidth]{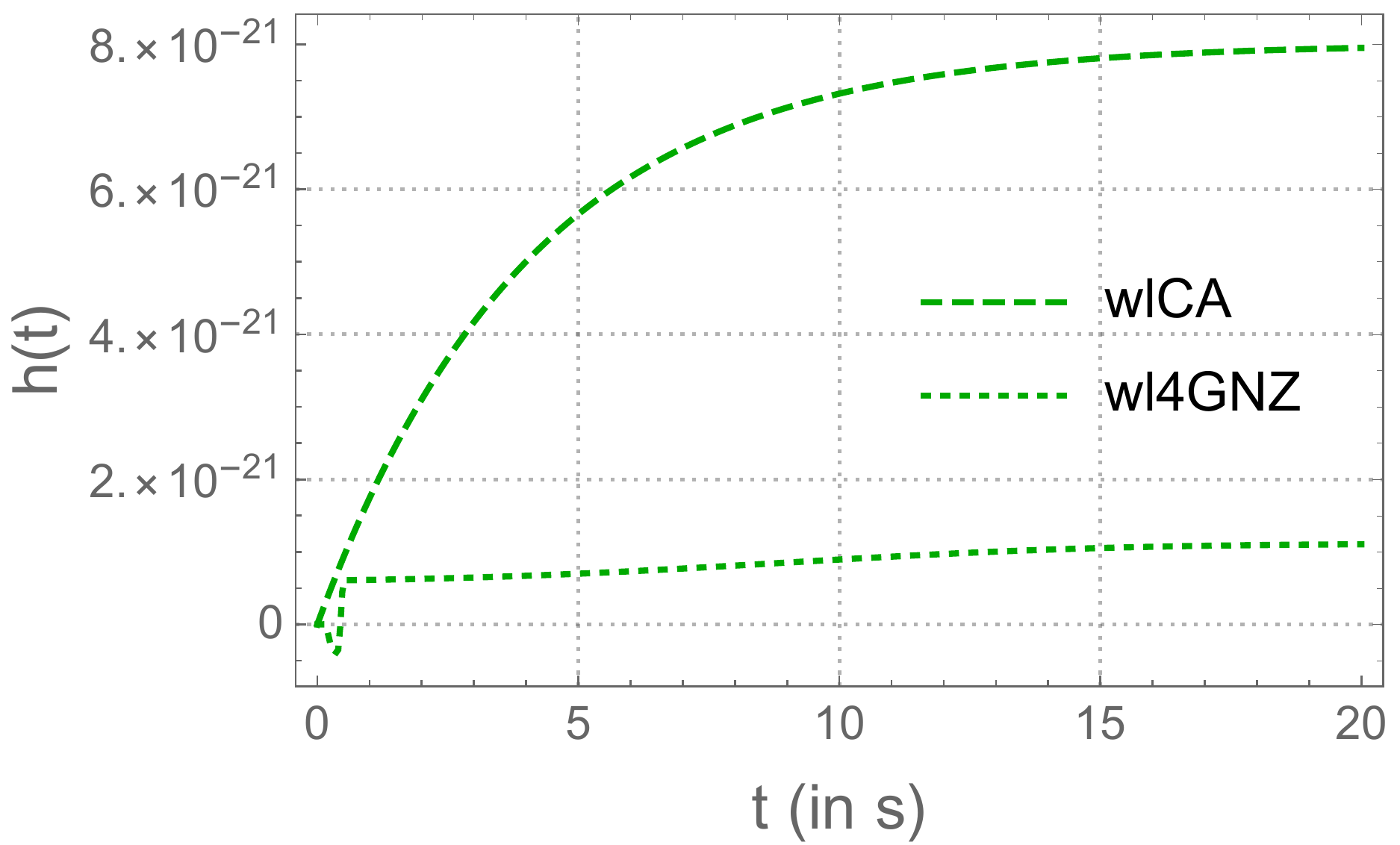}\label{fig:htp2}} 
        \subfloat[]{\includegraphics[width=0.49\textwidth]{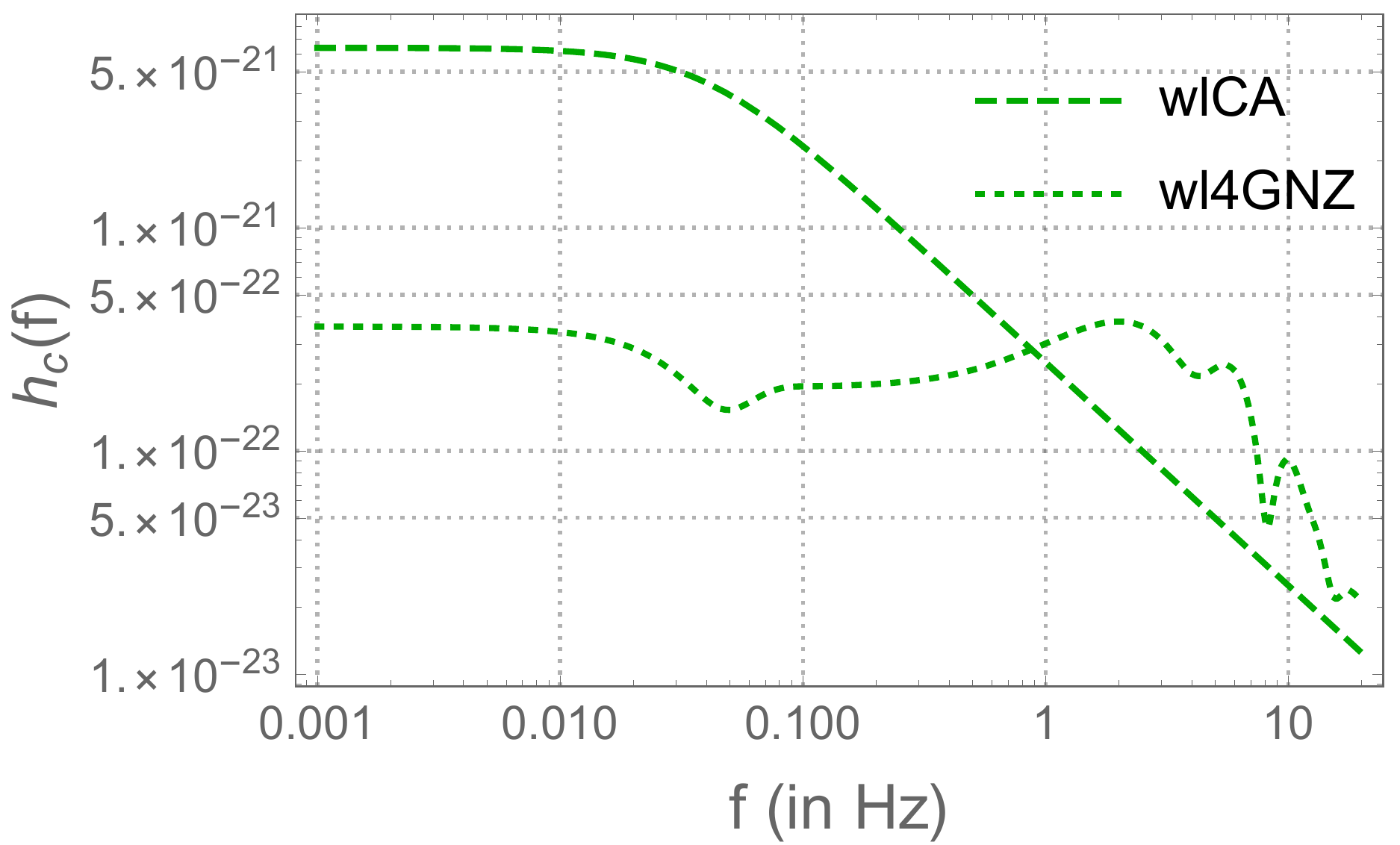}\label{fig:hcp2}}
        \end{center}
        \caption{\label{fig:ht_hc_plots}  \emph{Left panel :} Plots showing the dimensionless gravitational wave strain $h(t)$ for various models: (a) Solid blue line - accretion phase-three bumps model (Ac3G), Dot-dashed blue line - accretion phase-one bump model (Ac1G); (c) Dot-long dashed orange line - long accretion-three bumps model (LAc3G); (e) Dashed green line - whole luminosity-constant $\alpha$ model (wlCA), Dotted green line - whole luminosity-four bumps non-zero $\alpha$ model (wl4GNZ). \emph{Right panel :} Plots showing the characteristic gravitational wave strain $h_c(f)$ (in Hz) for various models: (b) Solid blue line - accretion phase-three bumps model (Ac3G), Dot-dashed blue line - accretion phase-one bump model (Ac1G); (d) Dot-long dashed orange line - long accretion-three bumps model (LAc3G); (f) Dashed green line - whole luminosity-constant $\alpha$ model (wlCA), Dotted green line - whole luminosity-four bumps non-zero $\alpha$ model (wl4GNZ). Note: The results for the models are color coded: The line color represents the luminosity for a particular phase from Figs.~\ref{fig:models2_3} and~\ref{fig:models1_4} and the line styles denote the corresponding anisotropy parameter from the same figures. In all the results shown above the distance to the source is taken to be $10$ kpc.
        }
\end{figure}

We now present results for the \mm\ strain in the time and frequency domain, $h(t)$ and $h_c$, for the phenomenological models of Sec. \ref{subsec:models} (Table \ref{tab: parameters}). They are shown in Fig. \ref{fig:ht_hc_plots}. 

For the accretion-only model, $h(t)$ reaches a plateau in less than one second, corresponding to the characteristic  duration of the episodes when the anisotropy parameter is non-zero (see Sec. \ref{subsec:template}).  Before the plateau, $h(t)$ can change sign due to the change of sign in $\alpha(t)$, as can be seen for the Ac3G model. Consistently with the upper bound, Eq. (\ref{eq:zfllim}), the late time, asymptotic value of $h(t)$ is $h \sim 10^{-22}$ m for the two more conservative scenarios (the value being the same for the Ac1G and Ac3G models is accidental, due to the choice of parameters, see Table \ref{tab: parameters}); it is largest -- approaching $\sim 10^{-20}$ -- for the most optimistic choice parameters, in the LAc3G model. 

For the long-term emission models, $h(t)$ evolves over the cooling time scale of several seconds. In the simplest case of constant anisotropy (wlCA model), it rises up smoothly to a value close to $\sim 10^{-20}$. The rise time reflects the decay time of the \n\ luminosity, and therefore it is directly comparable to the time profile of the detected \n\ signal. 
For the more realistic case of a time-evolving \ans\ parameter with a multi-second-wide Gaussian (wl4GNZ model), the evolution of $h(t)$ is the same as for the Ac3G model, with a weak late time rise that brings it to reach $\sim 10^{-21}$ at $t\sim 20$ s. 

In the figures for $h_c(f)$, one can check that the zero-frequency limit (ZFL) is indeed proportional  (by factor $1/\pi$) to the difference between the initial and final values of $h(t)$, Eq. (\ref{eq:zfl}) (Sec.~\ref{subsec:ZFLHFL}).  For the accretion only models, the ZFL is realized at $f\lesssim 0.1$ Hz. Above this value, one can start to see structures related to the time scale of the episode(s) of non-zero \ans.  We note how, for models where the \ans\ parameter does not change sign, $h_c$ is maximum at the ZFL. In the case of the Ac3G model, instead, where $\alpha(t)$ changes sign, the fast variability in $h(t)$ at $t \lesssim$0.5 s  causes $h_c$ to have a maximum at higher frequency, $f \sim 2$ Hz.   The same feature is seen, as expected, in the wl4GNZ model. 
For the two long-term evolution models, the ZFL is reached only at $f \lesssim 10^{-2}$ Hz, consistently with the longer evolution time scale. Note that these two models (wlCA and wl4GNZ) give comparable values of $h_c$ at $f\sim 1$ Hz, but are drastically different above and below this point, with the wlCA (wl4GNZ) model having significantly more power at lower (higher) frequency, as expected from the different time scales of $h(t)$ in the two models.   

\subsection{Detectability}
To estimate the potential for a memory signal to be detected at realistic \gw\ detectors, we compare the characteristic strains $h_c (f)$ with a typical characteristic detector noise amplitude, $h_n (f)$, averaged over the  source position and polarization angle.
Broadly, if $h_c (f) \gtrsim h_n(f)$ for a sufficiently wide frequency interval, the \mm\ can be considered likely to be detectable, although a detailed estimate of detectability is waveform specific, and would require a dedicated study. 

Following \cite{Li:2017mfz},  $h_n (f)$ can be expressed as
\be
\label{hnf}
h_n (f) = \frac{\sqrt{f S_n (f)}}{\langle F_+^2 (\theta,\phi,\psi)\rangle^{1/2}}\,,
\ee
where $S_n (f)$ is the detector's one-sided noise spectral density, in units of $s$ (${\rm Hz^{-1}}$) and $F_+ (\theta,\phi,\psi)$ is defined as the detector's beam pattern function (see~\cite{Sathyaprakash_2009, Moore_2014, Schmitz:2020syl} for details).  We choose the DECIGO as representative of the potential of future detectors at the Deci-Hz scale. DECIGO is planned to start, in prototype form, in the next decade \cite{Kawamura:2020pcg}. 


\begin{figure}[htb]
\begin{center}
    \includegraphics[width=0.90\textwidth,angle=0]{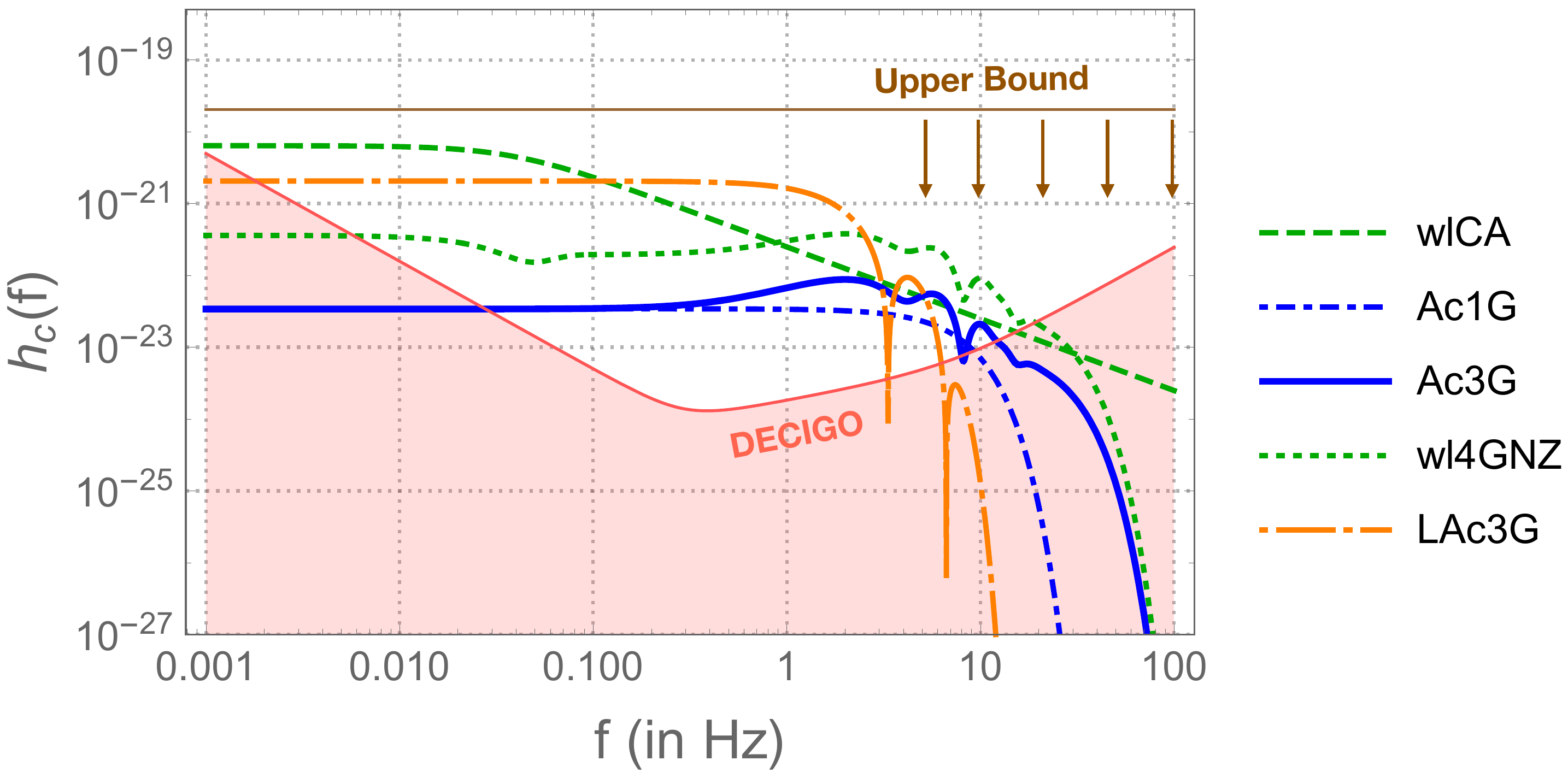}
\caption{\label{fig:hc_plot_decigo} Characteristic strain from the models along with a typical sky-averaged noise curve from DECIGO~\cite{Seto:2001qf,Yagi_2011,Sato_2017}. In all the results shown above (including the bound) the distance to the source is taken to be $10$ kpc. The upper bound, $h_c = 2.04 \times 10^{-20}$,~Eq. \eqref{eq:zfllim}, is shown as a horizontal line (with arrows).
}
\end{center}
\end{figure}

Fig.~\ref{fig:hc_plot_decigo} shows the sensitivity curve of DECIGO, compared with the results from the phenomenological models of Fig. \ref{fig:ht_hc_plots}, for a \sn\ at distance $r=10$ kpc.
It appears that, for all our models, the \mm\ signal is detectable up to  frequencies $f \sim 10$ Hz, and the zero frequency limit is observable (although only barely for the wlCA model). We note that for $f>  3$ Hz or so, the accretion only-models give the strongest signal, whereas for $f<0.1$ Hz $h_c$ is largest for the model with the strongest long-term \mm\ contribution (the wlCA model). Interestingly, in the intermediate range, $f \sim 0.1 - 3$ Hz, the LAc3G model would give the strongest signal, indicating that long accretion collapses (e.g., failed \sne) might be an especially promising target for Deci-Hz detectors. Considering the $1/r$ dependence of the \mm\ strain, from Fig.~\ref{fig:hc_plot_decigo} we estimate that a signal similar to the LAc3G model might be detectable at DECIGO for distances up to $r \sim 10$ Mpc or so. 

\begin{figure}
    \begin{center}
        \subfloat[]{\includegraphics[width=0.60\textwidth]{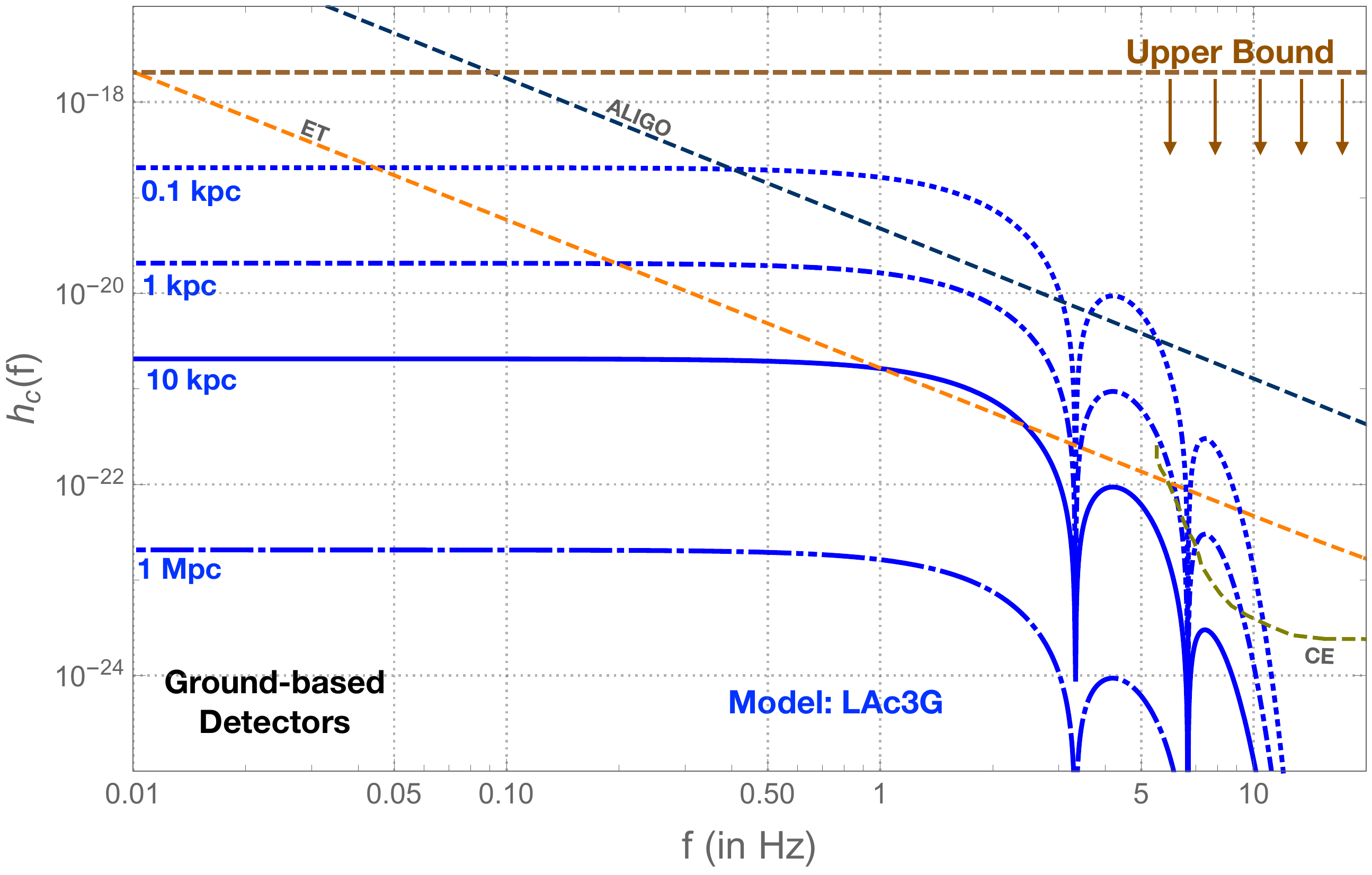}\label{fig:grndnf}} \\
        \subfloat[]{\includegraphics[width=0.60\textwidth]{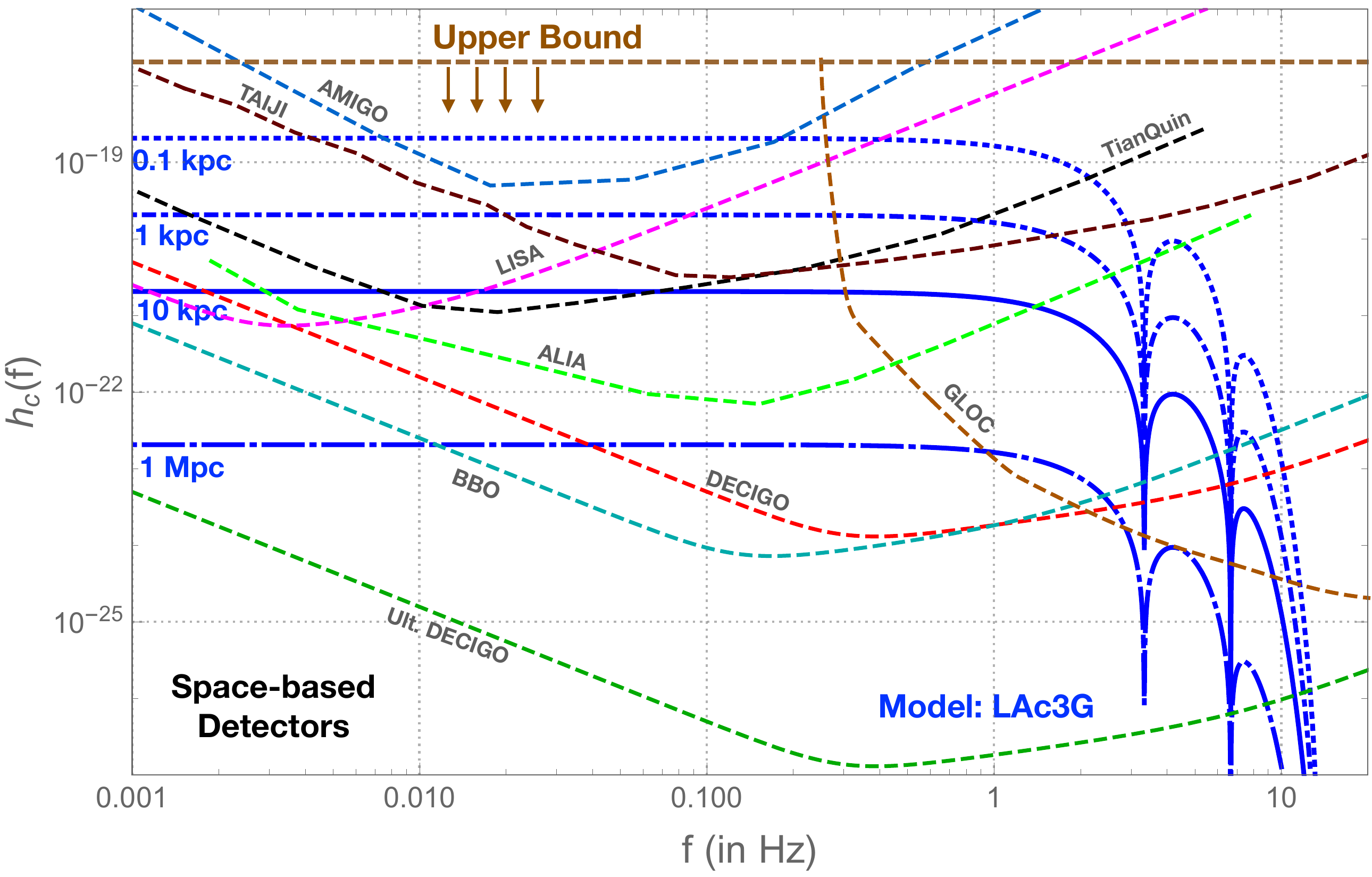}\label{fig:spcinf}} \\
        \subfloat[]{\includegraphics[width=0.60\textwidth]{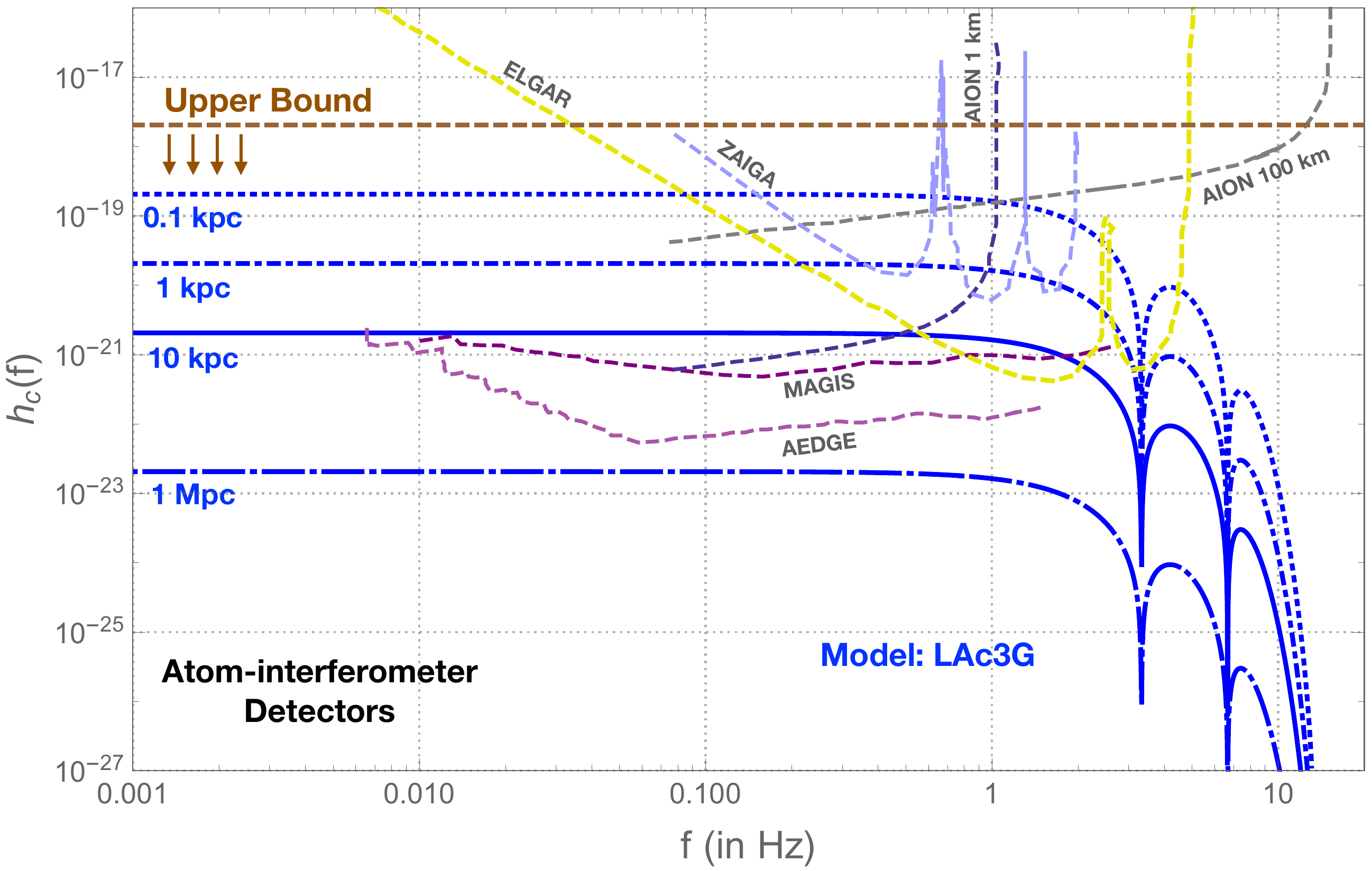}\label{fig:atminf}} 
    \end{center}
    \caption{\label{fig:lac3ginfs}  Characteristic strain of the supernova \n\ \mm\ from the LAc3G model for a \sn\  at distance $r=0.1, 1, 10, 10^3$ kpc, along with the upper bound from Eq. (\ref{eq:zfllim}) (shown for $r=0.1$ kpc). Detector sensitivity curves are shown, grouped in different panels as follows:   a) Ground-based detectors: ALIGO, ET, CE; b) Space-based detectors: LISA, DECIGO, Ultimate DECIGO, BBO, TAIJI, TianQuin, ALIA, GLOC, AMIGO ; c) Atom-interferometers: MAGIS, AEDGE, AION (1 km and 100 km), ZAIGA and ELGAR. See text for the full names of these projects and references. 
    }
\end{figure}

In Fig.~\ref{fig:lac3ginfs} we illustrate the potential of different detector concepts to observe the \n\ \mm\ from a \sn. %
A signal from an optimistic model, the LAc3G model, is shown for different distances to the \sn.
For comparison, we also plot the sensitivity curves for different next-generation detectors. 
We distinguish between the ground-based, space-based and atom-interferometer detectors. Ground-based interferometers (Fig.~\ref{fig:grndnf}) have limited performance at sub-Hz frequencies (due to seismic noise), and furthermore are not perfectly inelastic, a fact that would lead to the dissipation of memory effect signatures over time \cite{Favata:2010zu}. Among these, the Einstein Telescope (ET) \cite{Sathyaprakash_2009,Punturo_2010,Maggiore_2020} has the best potential, being able to observe the \mm\ at $f\sim 1-5$ Hz up to several kpc of distance to see the memory for a nearby ($\lesssim$ 1 kpc) supernova. In the same frequency range, the Advanced Laser Interferometer Gravitational-wave Observatory (ALIGO)~\cite{Sathyaprakash_2009,aligo2015} and Cosmic Explorer (CE)~\cite{Reitze:2019iox} could see a signature only for a near-Earth star like Betelgeuse ($r \sim$ 0.1 kpc).

The drawbacks suffered by ground based detectors are overcome by space-based interferometers (Fig.~\ref{fig:spcinf}), which have peak-performance at $f\sim 0.01 - 1 $ Hz. These 
are (in principle) completely inelastic, and therefore capable of preserving \mm\ signatures indefinitely. The most powerful proposed detectors of this type are the BBO~\cite{Yagi_2011},  and DECIGO;  for both of them the sensitivity to the signal extends up to $\sim 10$ Mpc. The most optimistic detector scenario, representative of the distant future potential, is Ultimate DECIGO \cite{Seto:2001qf,Yagi_2011,Sato_2017}, for which the distance of sensitivity exceeds 100 Mpc. Although less powerful, the Advanced Laser Interferometer Antenna (ALIA)~\cite{Bender_2013} and Laser Interferometer Space Antenna (LISA)~\cite{Sathyaprakash_2009,amaroseoane2017laser} should both be capable to detect a supernova at a typical galactic distance  ($r\sim 10$ kpc). 
Note that LISA will probe the zero-frequency limit, since it has peak sensitivity at the milli-Hz scale.

Interestingly, atom interferometry (Fig.~\ref{fig:atminf}) has recently emerged as an alternative to large scale traditional interferometers. 
After an initial stage on ground, most atom interferometer projects are envisioned to be in space, where their ultimate potential will be realized\footnote{An exception is the Zhaoshan long-baseline Atom Interferometer Gravitation Antenna (ZAIGA)~\cite{Zhan:2019quq}, which will be based underground.}. 
A space-based version of the Mid-band Atomic Gravitational Wave Interferometric Sensor (MAGIS) experiment \cite{PhysRevD.94.104022}, at the Km-length scale, might be sensitive to a galactic \sn. A prototype of MAGIS (with baseline of 100 m) is now approved for construction at the Fermi National Laboratory \cite{PhysRevD.94.104022, Graham:2017pmn,Coleman:2018ozp}.  A similar performance as MAGIS is expected for the Atomic Experiment for Dark matter and Gravity Exploration in space (AEDGE)~\cite{Bertoldi:2019tck}, which is being reviewed by the European Space Agency within its Voyage 2050 programme, the Atom Interferometer Observatory and Network (AION)~\cite{Badurina:2019hst} in its 1 km configuration, and the European Laboratory for Gravitation and Atom-interferometric Research (ELGAR)~\cite{Canuel:2019abg,Canuel:2020cxb}.

Other sensitivity curves shown in Fig.~\ref{fig:lac3ginfs}  are for TianQuin~\cite{Luo_2016}, the Gravitational-wave Lunar Observatory for Cosmology (GLOC)~\cite{jani2020gravitationalwave}, the Astrodynamical Middle-frequency Interferometric Gravitational wave Observatory (AMIGO)~\cite{Ni:2017bzv,Ni:2019nau} and the Zhaoshan long-baseline Atom Interferometer Gravitation Antenna (ZAIGA)~\cite{Zhan:2019quq}. 

\begin{figure}[htb]
\begin{center}
    \includegraphics[width=0.90\textwidth,angle=0]{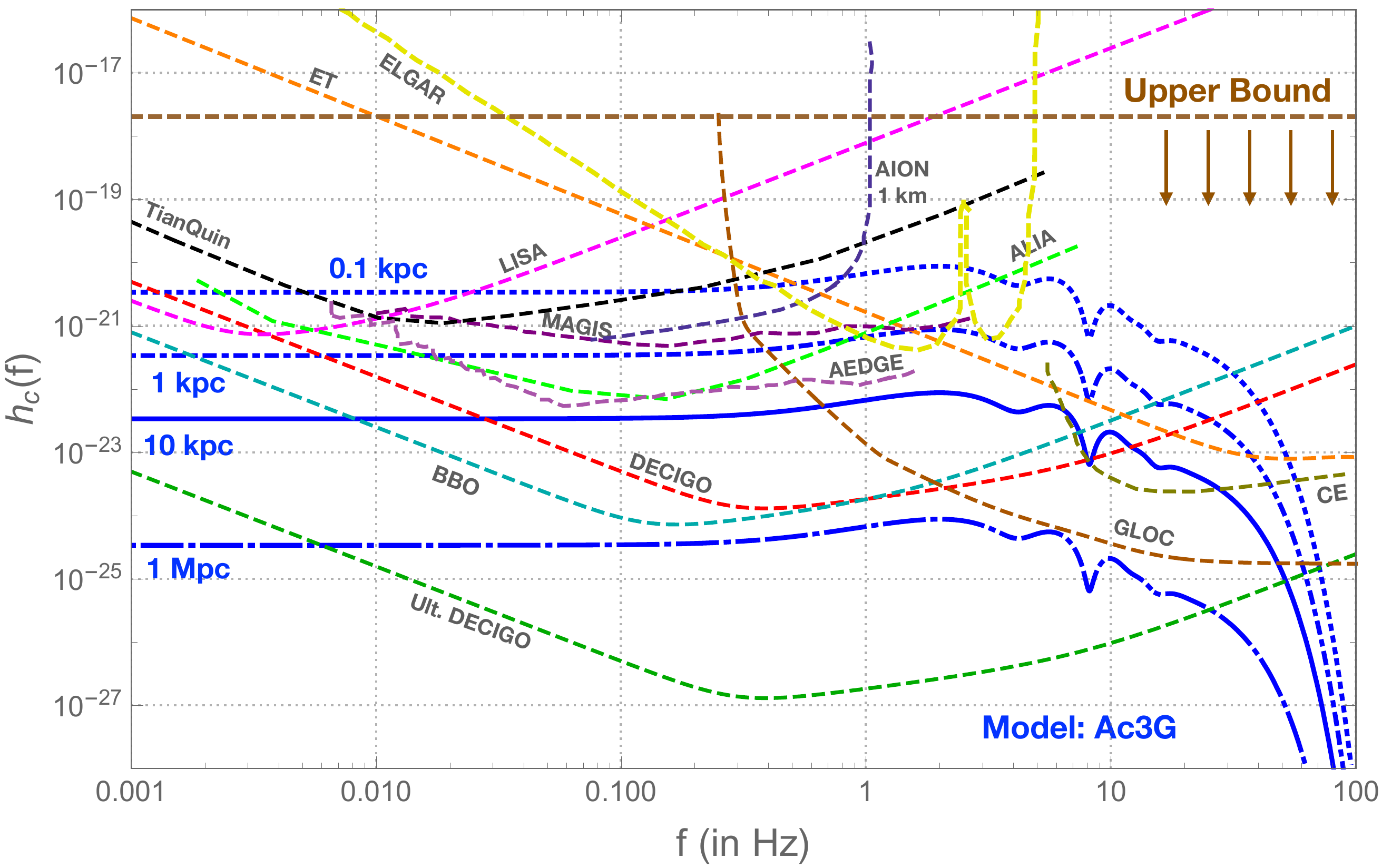}
\caption{\label{fig:final_plot} The same as Fig. \ref{fig:lac3ginfs}, for the Ac3G model. For simplicity, only sensitivity curves that intersect our theoretical predictions are plotted in a single panel. 
}
\end{center}
\end{figure}

A more conservative case for the observation of the \mm\ effect is given by the Ac3G model, as shown in Fig.~\ref{fig:final_plot}. For this model, the distance of sensitivity of each detector is reduced by about one order of magnitude, with MAGIS and DECIGO being limited to under 1 kpc and 1 Mpc respectively.

\section{Summary and discussion}
\label{sec:discussion}


We have presented a dedicated study of the phenomenology and detectability of the gravitational \mm\ effect due to the \n\ emission from a core collapse \sn. We place a theoretical upper bound on the strain, $h(t)$ (Eq.(\ref{eqn:uppernum})), and present phenomenological (analytical) description of the expected signal in time and frequency domain  (Eqs. (\ref{eq:mastereq}) and (\ref{eq:mastereq2})). This description reproduces the results of numerical simulations well, and its analytical form can be considered a generalization of earlier toy models of the \mm\ (e.g.,  \cite{Favata:2011qi}). 
We find that, depending on the type of scenario and on the choice of parameters, $h(t)$ and $h_c(f)$ vary between $\sim 10^{-23}$ and $\sim 10^{-20}$.  Due to the time-varying \ans\ parameter, $\alpha(t)$, the \mm\ signal is generally not straightforwardly related to the time profile of the \n\ luminosity profile. Rather, it could receive its dominant contribution from the second (or less)-long accretion phase, thus having substantial power at $f \gtrsim 1$ Hz.  

Naturally, large uncertainties affect our models, mostly due to our incomplete knowledge of the size and time evolution of the \ans\ parameter. While the accretion-only models are supported by state-of-the art numerical simulations of \sne, models with long-term \ans\ are necessarily more speculative, and have a character of plausibility. They suffer of uncertainties at the order of magnitude level. 
Let us also note that the \emph{total}, net \mm\ observed at Earth is uncertain due to the (uncertain) contribution of anisotropic matter ejection, which adds to the term due to \ns. The matter \mm\ should be subdominant at low frequency ($f \lesssim 1$ Hz), and become progressively more important in the high frequency regime, see, e.g., \cite{Kotake:2009rr,Vartanyan:2020nmt}. 

Our results confirm that, with the advent of Deci-Hz detectors, the observation of the gravitational \mm\ from a collapsing star will be realistic. Therefore,  the study of  \mm\ waveforms will become an important part of the multi-messenger approach to studying stellar core collapse, together with the observations of the \n\ burst at \n\ detectors (and, for near-Earth stars, of the pre-\sn\ \n\ emission,  \cite{ Patton:2017neq,Mukhopadhyay:2020ubs}), of gravitational waves at 10-100 Hz, and of the electromagnetic emission (if the collapse results in a \sn\ explosion).  
The best prospects for the \sn\ \mm\ detection are at the most powerful Deci-Hz detectors like BBO and DECIGO, for which the distance of sensitivity can reach $\sim 10$ Mpc for the most optimistic models, and $\sim 1$ Mpc for more conservative scenarios. Interestingly, these distances of sensitivity are comparable, or even larger than the typical reach of the largest realistic \n\ detector, a 0.3-1 Mt water Cherenkov detector like the approved HyperKamiokande \cite{Abe:2016ero}. This implies that a \mm\ detection might be an important early \sn\ alert for extragalactic \sne, that can be used for astronomy and \n\ follow ups.  

The most immediate physics potential of the observation of a \n\ \mm\ signal is the possibility to probe the \ans\ parameter, $\alpha(t)$. From the \mm\ alone -- especially if probed over a wide range of frequencies, through the interplay of multiple interferometers   --  it might be possible to distinguish between drastically different scenarios. In particular, a fast-rising, fast-plateauing \mm\ signal, with frequency peak at or beyond 1 Hz would be an indication of an accretion-only scenario, whereas a slower rising (multi-second time scale), lower frequency (peak at  $f \lesssim 1$ Hz) \mm\ strain will indicate the presence of a long-term \ans. 

For a galactic \sn, the interplay with a high statistics observation of the \n\ burst will open the possibility of precision measurement, because $\alpha(t)$ can be extracted by comparing the \mm\ data with the \emph{measured total \n\ luminosity} (for which high sensitivity to the non-electron \n\ flavor is important, a fact that strengthens the motivations of efforts on this front). The measured anisotropy parameter could carry imprints of the hydrodynamics happening during the core-collapse, such as SASI, turbulence and chaotic dynamics, thus providing  an important test of numerical models of core collapse.
A detailed comparison of the \mm\ features with SASI signatures observed in \ns\ and/or in gravitational waves at $\sim 100$ Hz may allow to pinpoint important events in the dynamics of the collapsed star, such as a change in the plane of the spiral SASI. 
Using the \n\ burst data, it might also be possible to disentangle the matter contribution to the \mm\ from the \n\ one,  thus offering a new handle on the phenomena that contribute to it. 

Outside the field of multimessenger astronomy, the theme of \ns\ as sources of the gravitational \mm\ is worth further exploration for its significance in general relativity. 
In particular, beyond the 
linear memory studied in this work, it would be worth to ask if neutrino emissions would play any significant role into the non-linear memory effect \cite{Christodoulou:1991cr, Blanchet:1992br}, which accounts for the contribution to the memory from the gravitational wave itself. It has been shown  \cite{Thorne:1992sdb} that the non-linear memory can be described by a linear memory in which the sources are the individual radiated gravitons; therefore, one might wonder if an analogous effect occurs from the emitted neutrinos. Even more recently, it has been argued \cite{Favata:2010zu} that the non-linear memory has a large contribution in the gravitational waveform which enter at leading order in a post-Newtonian expansion. The reason of this large contribution being the hereditary nature of the memory, which can be build from long times for some long lived asymmetries of the source. Another exciting aspect of the non-linear memory, is its connection to the group of symmetries for asymptotically flat space-time metrics \cite{Strominger:2013jfa} which is directly related to the study of the vacuum in quantum gravity and the infrared structure of gravity. Along the same lines, it has been shown that the non-linear memory is just equivalent to Weinberg’s theorem  for soft graviton production \cite{Strominger:2014pwa}. All these new developments motivate a study of the non-linear \mm\ in the novel context of neutrino emission. 

In closing, the gravitational \mm\ could be the next major prediction of general relativity to receive an impressive experimental confirmation, in a not-too-distant future. It is interesting that such a first observation will be directly linked to another exciting event, the detection of a \n\ burst from a galactic (or near-galactic) supernova. Learning about gravity from neutrinos, and vice-versa, will be a new and fascinating development in multimessenger astronomy. 

\newpage
\begin{table}
\centering
\begin{tabular}{||M{29mm}||M{25mm}||M{30mm}|| M{35mm} ||}
\hline
\vspace{0.2cm} Curve \vspace{0.2cm} & \vspace{0.2cm} $L_{\nu} (t)$ Parameters \vspace{0.2cm} &\vspace{0.2cm} $\alpha (t)$ Parameters \vspace{0.2cm} & \vspace{0.2cm} Effective Parameters \vspace{0.2cm} 
\\
\hline
Analytical (Blue Dashed) & $\lambda = 3.16 \times 10^{52}$ ergs/s, $\beta = 2.25 \times 10^{52}$ ergs/s, $\chi = 2.24\ s^{-1}$ & $N = 3$, $\kappa=0.0$, $\xi_1 = -0.01$, $\xi_2 = -0.015$, $\xi_3 = 0.02$, $\gamma_1=0.205$ s, $\gamma_2=0.325$ s $\gamma_3=0.45$ s, $\sigma_1=0.04/\sqrt{2}$ s, $\sigma_2=0.04/3\sqrt{2}$ s, $\sigma_3=0.04/\sqrt{2}$ s & \vspace{0.8cm} $h_{11}=3.12 \times 10^{-23}$, $h_{12}= -1.03 \times 10^{-23}$, $h_{13} =-2.69 \times 10^{-23}$, $\rho_{1} = 25$ Hz, $\rho_{2} = 75$ Hz, $\rho_{3} = 25$ Hz, $\tau_{11}= 0.4482$ s, $\tau_{12}= 0.3248$ s, $\tau_{13}= 0.2032$ s, $h_{21}= 1.20 \times 10^{-22}$, $h_{22}= -2.99 \times 10^{-23}$, $h_{23}= -5.98 \times 10^{-23}$, $\tau_{21}= 0.45$ s, $\tau_{22}= 0.325$ s, $\tau_{23}= 0.205$ s, $h_{3}=0.0$ \vspace{0.8cm}
\\
\hline
Phenomenological (Black Dot-dashed) & $\lambda =5.97 \times 10^{51}$ ergs/s, $\beta = 2.25 \times 10^{52}$ ergs/s, $\chi = 2.24\ s^{-1}$ & $N = 3$, $\kappa=0.0$, $\xi_1 = 0.01$, $\xi_2 = -0.04$, $\xi_3 = 0.078$, $\gamma_1=0.100$ s, $\gamma_2=0.250$ s $\gamma_3=0.450$ s, $\sigma_1=0.04/\sqrt{2}$ s, $\sigma_2=0.04/\sqrt{2}$ s, $\sigma_3=0.04/\sqrt{2}$ s & \vspace{0.8cm} $h_{11}=1.22 \times 10^{-22}$, $h_{12}= -9.74 \times 10^{-23}$, $h_{13} =3.41 \times 10^{-23}$, $\rho_{1} = 25$ Hz, $\rho_{2} = 25$ Hz, $\rho_{3} = 25$ Hz, $\tau_{11}= 0.4482$ s, $\tau_{12}= 0.2482$ s, $\tau_{13}= 0.0982$ s, $h_{21}= 8.81 \times 10^{-23}$, $h_{22}= -4.52 \times 10^{-23}$, $h_{23}= 1.13 \times 10^{-23}$, $\tau_{21}= 0.45$ s, $\tau_{22}= 0.25$ s, $\tau_{23}= 0.10$ s, $h_{3}=0.0$ \vspace{0.8cm}
\\
\hline
\end{tabular}
\caption{\label{tab: fit_parameters}  Table showing the parameters relevant for luminosity $L_\nu (t)$ and anisotropy parameter $\alpha (t)$ along with the respective effective parameters as defined in~\eqref{eq:htformb}, corresponding to the different curves in Fig.~\ref{fig:num_ana_compare}. Note: The normalization for the luminosity is different for the two curves, since the analytical curve is just a curve superimposed on the raw data. 
}
\end{table}


\begin{table}
\centering
\begin{tabular}{||M{18mm}||M{14mm}||M{25mm}||M{32mm}|| M{40mm} ||}
\hline
Model Name & Acronym & $L_{\nu} (t)$ Parameters & $\alpha (t)$ Parameters & Effective Parameters \\
\hline
Accretion phase - three bumps & Ac3G (N=3) & $\lambda= 5.97 \times 10^{51}$ ergs/s, $\beta = 2.25 \times 10^{52}$ ergs/s, $\chi = 2.24\ s^{-1}$ & $\kappa=0.0$, $\xi_1 = -0.01$, $\xi_2 = -0.015$, $\xi_3 = 0.04$, $\gamma_1=0.205$ s, $\gamma_2=0.325$ s $\gamma_3=0.45$ s, $\sigma_1=0.04/\sqrt{2}$ s, $\sigma_2=0.04/3\sqrt{2}$ s, $\sigma_3=0.04/\sqrt{2}$ s & $h_{11}=6.23 \times 10^{-23}$, $h_{12}= -1.03 \times 10^{-23}$, $h_{13} = -2.69 \times 10^{-23}$, $\rho_{1} = 25$ Hz, $\rho_{2} = 75$ Hz, $\rho_{3} = 25$ Hz, $\tau_{11}= 0.4482$ s, $\tau_{12}= 0.3248$ s, $\tau_{13}= 0.2032$ s, $h_{21}= 4.52 \times 10^{-23}$, $h_{22}= -5.65 \times 10^{-24}$, $h_{23}= -1.13 \times 10^{-23}$, $\tau_{21}= 0.45$ s, $\tau_{22}= 0.325$ s, $\tau_{23}= 0.205$ s, $h_{3}=0.0$\\
\hline
Accretion phase - one bump & Ac1G (N=1) & $\lambda= 5.97 \times 10^{51}$ ergs/s, $\beta = 2.25 \times 10^{52}$ ergs/s, $\chi = 2.24\ s^{-1}$ & $\kappa = 0.0$, $\xi_1 = 0.02$, $\gamma_1=0.45$ s, $\sigma_1=0.04/\sqrt{2}$ s &  $h_{11}= 3.12 \times 10^{-23}$, $\rho_{1} = 25.0$ Hz, $\tau_{11}= 0.4482$ s, $h_{21}= 2.26 \times 10^{-23}$, $\rho_{2} = 25.0$ Hz, $\tau_{21}= 0.45$ s, $h_{3}=0.0$\\
\hline
Long Accretion phase - three bumps & LAc3G (N=3) & $\lambda= 1.0 \times 10^{53}$ ergs/s, $\beta = 0.0$, $\chi = 0.0$ & $\kappa=0.0$, $\xi_1 =  \xi_2 = \xi_3 =0.023$, $\gamma_1=\gamma_2=\gamma_3=0.35$ s, $\sigma_1=\sigma_2=\sigma_3=0.07$ s & $h_{1i}=0.0$, $\rho_{1} = \rho_{2} = \rho_{3} = 10.10$ Hz, $\tau_{21}= 0.35$ s, $\tau_{22}= 0.45$ s, $\tau_{23}= 0.55$ s, $h_{21}= h_{22}=h_{23}= 1.08 \times 10^{-21}$, $h_{3}=0.0$\\
\hline
Whole Luminosity - constant alpha & wlCA (N=0) & $\lambda= 0.0$, $\beta = 7.35 \times 10^{52}$ ergs/s, $\chi = 0.245\ s^{-1}$ & $\kappa = 0.005$ & $h_{1i}=0.0$, $h_{2i}=0.0$, $h_3= 2.67 \times 10^{-67}$\\
\hline
Whole Luminosity - four bumps non-zero alpha & wl4GNZ (N=4) & $\lambda= 0.0$, $\beta = 7.35 \times 10^{52}$ ergs/s, $\chi = 0.245\ s^{-1}$ & $\kappa=0.0$, $\xi_1 = -0.01$, $\xi_2 = -0.015$, $\xi_3 = 0.04$, $\xi_4 = 0.002$, $\gamma_1=0.205$ s, $\gamma_2=0.325$ s $\gamma_3=0.45$ s, $\gamma_4=15.0$ s, $\sigma_1=0.04/\sqrt{2}$ s, $\sigma_2=0.04/3\sqrt{2}$ s, $\sigma_3=0.04/\sqrt{2}$ s, $\sigma_4=5.0$ s &  $h_{11}=4.98 \times 10^{-22}$, $h_{12}= -6.42 \times 10^{-23}$, $h_{13} = -1.32 \times 10^{-22}$, $h_{14}= 2.64 \times 10^{-22}$, $\rho_{1} = 25$ Hz, $\rho_{2} = 75$ Hz, $\rho_{3} = 25$ Hz,  $\rho_{4} = 0.1414$ Hz, $\tau_{11}= 0.4498$ s, $\tau_{12}= 0.3250$ s, $\tau_{13}= 0.2048$ s, $\tau_{14}= 8.875$ s, $h_{2i}=0.0$, $h_{3}=0.0$\\
\hline
\end{tabular}
\caption{\label{tab: parameters} Table showing the parameters relevant for the luminosity $L_{\nu} (t)$ and anisotropy parameter $\alpha (t)$ for different analytical models constructed along with the respective effective parameters as defined in~\eqref{eq:htformb}. Here, N is the number of Gaussians in $\alpha(t)$.
}
\end{table}

\newpage
\appendix

\section{Formalism Addendum}
\label{sec:appendix A}
In this appendix we would like to collect some details to complement section \ref{sec:formalism}
\subsection{The weak field equations}
\label{subsec:feild_eqns}
Consider a metric $g_{\mu \nu}$ nearly flat\footnote{ We use signature $\eta_{\mu \nu} = diag(-1,1,1,1)$.} as given in Eq.~\eqref{metric},
\be
g_{\mu \nu} = \eta_{\mu \nu} + h_{\mu \nu}.
\ee
i.e, $h_{\mu \nu}$ is a small perturbation and we  will only keep terms up to first order in $h_{\mu \nu}$. The Ricci tensor is defined as,
\be
\label{ricciten}
R_{\mu \nu} = \pd_\nu \Gamma^\lambda_{\lambda \mu} - \pd_\lambda \Gamma^\lambda_{\nu \mu},
\ee
where, 
\be
\label{christoffel}
\Gamma^\lambda_{\mu \nu} = \half \eta^{\lambda \rho} \Big( \pd_\mu h_{\rho \nu} + \pd_\nu h_{\rho \mu} - \pd_\rho h_{\mu \nu} + \mathcal{O}(h^2) \Big).
\ee
Using Eq. \ref{ricciten} and \ref{christoffel}, we have;
\be
\label{riccieqn}
R_{\mu \nu} \approx R^{(1)}_{\mu \nu} \equiv \half \Big( \Box^2 h_{\mu \nu} -\pd_\lambda \pd_\mu h^{\lambda}_\nu - \pd_\lambda \pd_\nu h^{\lambda}_{\mu} + \pd_\lambda \pd_\nu h^{\lambda}_\lambda \Big).
\ee
Recall the Einstien's field equation is defined as,
\be
\label{einseqn}
R_{\mu \nu} - \half R g_{\mu \nu} = -8 \pi G T_{\mu \nu},
\ee
(Note that, $R = 8 \pi G T^\mu_\mu$).\\
Using Eq. \ref{riccieqn} in Eq. \ref{einseqn} after some massaging we obtain the following field equation,
\be
\label{fieldequations}
\Box^2 h_{\mu \nu} -\pd_\lambda \pd_\mu h^{\lambda}_\nu - \pd_\lambda \pd_\nu h^{\lambda}_{\mu} + \pd_\lambda \pd_\nu h^{\lambda}_\lambda = - 16 \pi G S_{\mu \nu}
\ee
where $S_{\mu \nu}$ is defined in terms of the conventional stress-energy tensor $T_{\mu\nu}$ as,
\be
\label{stresstensor}
S_{\mu \nu} = T_{\mu \nu} - \half \eta_{\mu \nu}T^\lambda_\lambda.
\ee
As we mention in section \ref{sec:formalism}, the  field equation above is gauge invariant. We choose a particular gauge
\be
\label{gaugecond}
g^{\mu \nu} \Gamma^\lambda_{\mu \nu} = 0.
\ee
which from \eqref{christoffel} implies the following condition on the metric perturbation, 
\be
\label{gaugeeqn}
\pd_\alpha h^\alpha_\lambda = \half \pd_\lambda h^\mu_\mu,
\ee
This finally gives us,
\be
\label{mainfieldeqn}
\Box^2 h_{\mu \nu} = -16 \pi G S_{\mu \nu}
\ee
One can write a solution by using the retarded Green's function and explicitly take,
\be
\label{greensol}
h_{\mu \nu} = 4G \int d^3 \vec{x}^{\,'} \Big( \frac{S_{\mu \nu}(\vec{x}^{\,'},t-|\vec{x}-\vec{x}^{\,'}|)}{|\vec{x}-\vec{x}^{\,'}|} \Big).
\ee
Following \cite{Epstein:1978dv}, we use the following ansatz for the sources,
\be
\label{source}
S^{ij}(t,x) = n^i n^j r^{-2} \sigma(t-r)f(\Omega,t-r),
\ee
where, $\vec{n} = x/r$, $r=|x|$. This stress-tensor represents a point source that releases matter at $x=0$ at the speed of light, with $\sigma(t)$ the rate of energy loss and $f(\Omega,t)$ the angular distribution of emission and hence it satisfies $f (\Omega,t)\geq 0$ and $\int f(\Omega,t)d\Omega = 1$. A convenient way of writing the source ansatz \ref{source} is,
\be
\label{sourcedelta}
S^{ij}(t,x) = n^i n^j r^{-2} \int_{-\infty}^{\infty} f(\Omega^{\prime},t^{\prime}) \sigma(t^{\prime})\delta (t-t^{\prime}-r) dt^{\prime}.
\ee
In order to fix the residual gauge freedom (see for example \cite{Misner:1974qy}), we should project the source stress-tensor into its transverse-traceless component, which we denote by $(n^i n^j)_{TT}$ and write explicitly in the wave form as \footnote{We delay an explicit computation of $(n^i n^j)_{TT}$ to appendix \ref{appsec:nxnxtt} below.},
\ba
\label{epseq15}
&&h^{ij}_{TT}(t,x)=\nonumber\\ && \qquad4 \int_{-\infty}^{\infty} \int_{4 \pi} \int_{0}^{\infty} \frac{(n^i n^j)_{TT} f(\Omega^{\prime},t^{\prime}) \sigma(t^{\prime})}{|\vec{x}-\vec{x^{\prime}}|}  \delta (t-|\vec{x}-\vec{x^{\prime}}|-t^{\prime}-r^{\prime}) dr^{\prime} d \Omega^{\prime} dt^{\prime}.    
\ea
We use the identity $|x-x^{\prime}|^2 = r^2 + r^{\prime 2} - 2r r^{\prime} \cos{\theta}$ and perform the integration with respect to $r^{\prime}$ to get rid of the delta function, leading us to,
\be
\label{epseq16}
h^{ij}_{TT}(t,x) = 4 G \int_{-\infty}^{t-r} \int_{4 \pi} \frac{(n^i n^j)_{TT} f(\Omega^{\prime},t^{\prime}) \sigma(t^{\prime})}{t-t^{\prime}-r \cos{\theta}} d \Omega^{\prime} dt^{\prime}.
\ee

\subsection{From gravitation to neutrino physics}
\label{subsec: neutrino_formula}

For the derivation of the equations in this appendix, we have followed \cite{janka_muller, Li:2017mfz, Ott:2008wt, Burrows:1995bb, Kotake:2012iv}.  
\\
The observer is situated at a distance $r=|x|\rightarrow \infty$ from the source, \emph{i.e.}, very far away from the event, and sees the radiation from the source at a time $t$ which was emitted at time $t^{\prime} = t-r/c$. We are interested in the gravitation wave created by the neutrino pulse. By defining the direction dependent neutrino luminosity as\footnote{In other words, the energy radiated at time $t$ per unit of time and per unit of solid angle into the direction $\Omega^{\prime}$.} $\frac{dL_\nu (\Omega^{\prime},t^{\prime})}{d \Omega^{\prime}}=f(\Omega^{\prime},t^{\prime}) \sigma(t^{\prime}) $,
 we can now rewrite ~\eqref{epseq16} in this approximation as,
\be
\label{eqn:httform2}
h^{ij}_{TT}(t,x) = \frac{4G}{r c^4} \int_{-\infty}^{t-r/c} dt^{\prime} \int_{4 \pi} \frac{(n^i n^j)_{TT}}{1 - \cos{\theta}} \frac{dL_\nu (\Omega^{\prime},t^{\prime})}{d \Omega^{\prime}} d \Omega^{\prime},
\ee
The wave $h^{ij}_{TT}(t,x)$ can be either `$+$' or `$\times$' polarized. We denote the $+$ polarization as, $h^{xx}_{TT} = -h^{yy}_{TT} = -h^{+}_{TT}$. 

By defining the \emph{anisotropy parameter} $\alpha(t)$ in the x-direction as given by,
\be
\label{anisotropy}
\alpha(t) = \frac{1}{L_{\nu}(t)} \int_{4 \pi} d \Omega^{\prime}\ \frac{(n^x n^x)_{TT}}{1 - \cos{\theta}}\ \frac{dL_{\nu} (\Omega^{\prime},t)}{d \Omega^{\prime}}\,,
\ee
we can conveniently rewrite the wave form as,
\be
h^{xx}_{TT} = \frac{2G}{r c^4} \int_{-\infty}^{t-r/c} dt^{\prime} L_{\nu}(t^{\prime}) \alpha(t^{\prime}),
\ee
where the total neutrino luminosity is given by,
\be
\label{luminosity}
L_{\nu}(t) = \int_{4 \pi} d \Omega^{\prime}\ \frac{dL_{\nu} (\Omega^{\prime},t)}{d \Omega^{\prime}}.
\ee

\section{Calculation of transverse-traceless amplitude}\label{appsec:nxnxtt}

There is still an important piece $(n^i n^j)_{TT}$ at equations \eqref{eqn:httform2} and \eqref{anisotropy} that requires some work since it contains the angular dependence for the integrand of the wave form. In this section we will compute the transverse traceless component of the wave form.
\begin{figure}
\begin{center}
    \includegraphics[width=0.65\textwidth,angle=0]{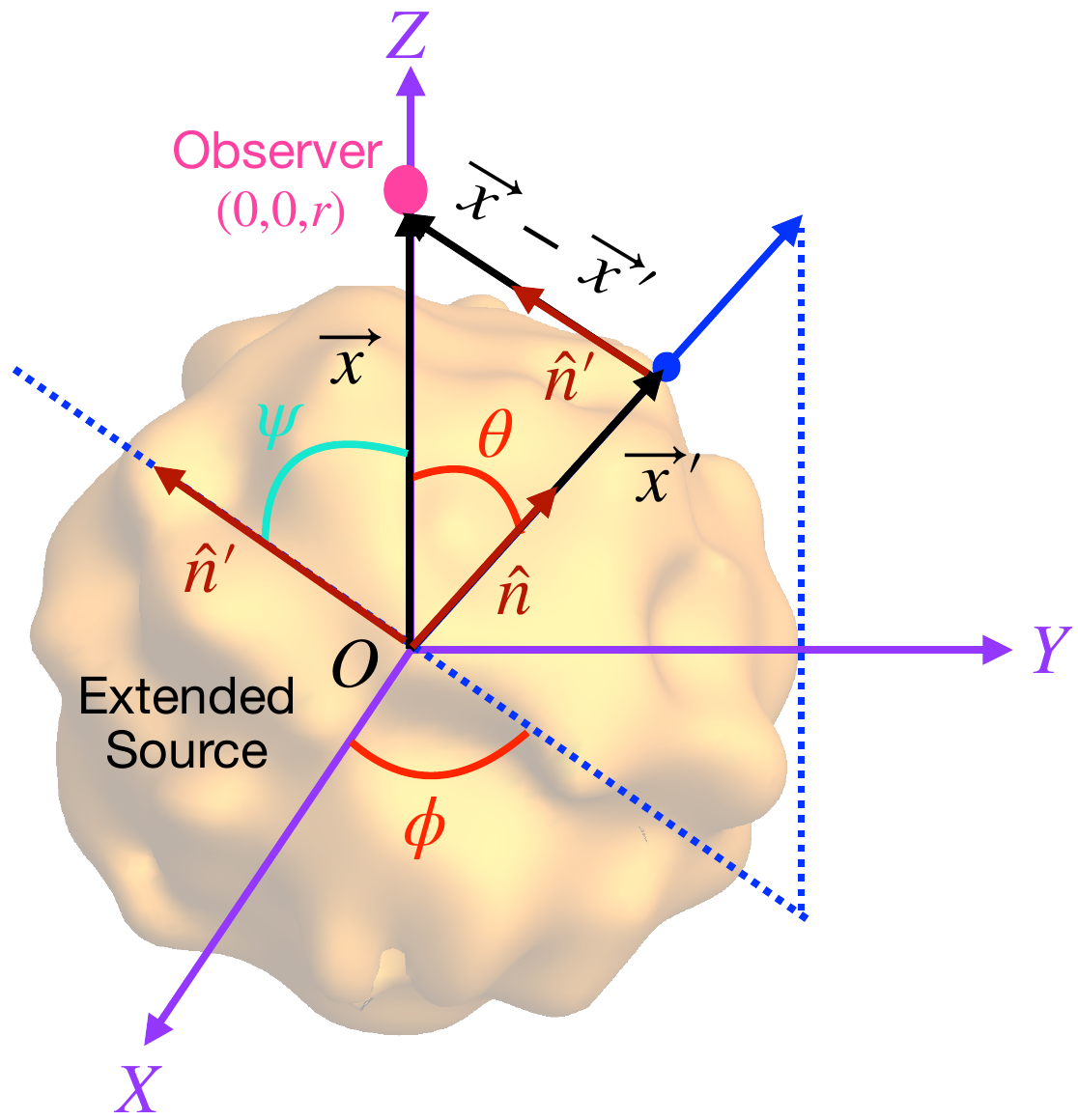}
\caption{\label{appfig:setup2} Setup to show the different vectors involved. Same as Fig.~\ref{fig:setup1} (including the colors used) but only the observer's coordinate system is shown along with the extended source. The blue dot is a point on the surface of the extended source and the corresponding position vector is shown as a blue arrow.
}
\end{center}
\end{figure}

From Fig.~\ref{appfig:setup2} one can read the vectors, 

\begin{equation}
\begin{aligned}
\hat{n}&= \frac{\vec{x^{\prime}}}{|\vec{x^{\prime}}|}\\
\hat{n}^{\prime}&= \frac{\vec{x} - \vec{x}^{\prime}}{|\vec{x} - \vec{x}^{\prime}|}\\
\end{aligned}
\label{vectors}
\end{equation}
or in components,
\begin{equation}
\begin{aligned}
n_x&= \sin{\theta} \cos{\phi}\\
n_y&= \sin{\theta} \sin{\phi}\\
n_z&= \cos{\theta}
\end{aligned}
\label{ncomp}
\end{equation}

\begin{equation}
\begin{aligned}
n^{\prime}_x&= - \sin{\psi} \cos{\phi}\\
n^{\prime}_y&= - \sin{\psi} \sin{\phi}\\
n^{\prime}_z&= \cos{\psi}
\end{aligned}
\label{npcomp}
\end{equation}
From Eq. \ref{ncomp}, \ref{npcomp}, it is easy to show that,
\be
\label{ndotnp}
\hat{n} . \hat{n}^{\prime} = \cos{(\theta + \psi)}.
\ee
Also, from the figure (Fig. \ref{appfig:setup2}) we see that $\vec{x}.\hat{n}^{\prime} = r \cos{\psi}.$
The operator that projects into the transverse-traceless component is given in the standard way,
\be
P^i_j = \delta^i_j - n^{\prime i}n^{\prime}_j.
\label{projop}
\ee
More explicitly, the transverse-traceless (TT) part of a given tensor $T^{kl}$ is given by,
\be
T^{ij}_{TT} = \Big( P^i_k P^j_l - \frac{1}{2}P^{ij}P_{kl} \Big) T^{kl}.
\label{ttop}
\ee
Let us now check that our operator in Eq. \ref{projop} does what it is supposed to, namely project out the transverse-traceless part of a tensor. First, we check that the component \ref{ttop} is transverse to the direction of wave propagation. In other words, $T^{ij}_{TT}$ should be perpendicular (transverse) to $\np$ and therefore its dot product with $\np$ should vanish,
\begin{equation*}
    \np_i T^{ij}_{TT} = (\np_k - \np_i \np^i \np_k)(\delta^j_l - \np^j \np_l)T^{kl}\\ - \frac{1}{2}(\np_j - \np_i \np^i \np_j)(\delta^k_l - \np^k\np_l)T^{kl} = 0,
\end{equation*}
which is what we required. Next, we want to show that the tensor component  $T^{ij}_{TT}$ is also traceless by showing $\delta^i_j T^{ij}_{TT} = 0$. Once again substituting the expressions from Eq.~\ref{projop}, \ref{ttop} and doing the appropriate contractions give us,
\begin{align*}
    \delta^i_j T^{ij}_{TT} &= (\delta^j_k - \np^j \np_k)(\delta^j_l - \np^j \np_l)T^{kl} - \frac{1}{2}(3-1)(\delta^k_l - \np^k \np_l)T^{kl} \\ &= (\delta^k_l - \np^k \np_l)T^{kl} - (\delta^k_l - \np^k \np_l)T^{kl} = 0,
\end{align*}
which is what we intended to show. 
Now we want to evaluate the expression $(n^x n^x)_{TT}$,
\begin{equation}
(n^x n^x)_{TT} = \Big( P^x_k P^x_l - \frac{1}{2}P^{xx}P_{kl} \Big) n^k n^l 
= \Big(P^x_k n^k n^l P^x_l - \frac{1}{2}P^{xx}P_{kl}  n^k n^l \Big)
\label{nxnx}
\end{equation}

Let us first calculate each term separately appearing in the above expression.
\begin{equation*}
P^x_k n^k =  (\delta^x_k - \np^x \np_k)n^k 
= n^x - \np^x \cos{(\theta + \psi)} 
= \sin{\theta} \cos{\phi} + \sin{\psi} \cos{\phi} \cos{(\theta + \psi)}
\end{equation*}
\be
P^x_k n^k = \cos{\psi} \cos{\phi} \sin{(\theta + \psi)}
\label{projx}
\ee
In the above computations we have used the expressions in Eq.~\ref{ncomp}, \ref{npcomp}, \ref{ndotnp}, \ref{projop}. Similarly one can find the following projections,
\be
P^y_k n^k = \cos{\psi} \sin{\phi} \sin{(\theta + \psi)}
\label{projy}
\ee
\be
P^z_k n^k = \sin{\psi} \sin{(\theta + \psi)}
\label{projz}
\ee
Next we have,
\begin{equation*}
P_{kl} n^k n^l = (\delta_{kl} - \np_k \np_l)n_k n_l 
= 1- \cos^2{(\theta + \psi)},
\end{equation*}
\be
P_{kl} n^k n^l = \sin^2{(\theta + \psi)},
\label{projsum}
\ee
and finally,
\be
P^{xx} = (\delta^x_x - \np^x \np_x) = 1 - \sin^2{\psi}\cos^2{\phi}.
\label{pxx}
\ee
So now we have all the terms required for the evaluation of $(n^x n^x)_{TT}$. Substituting Eq.~\ref{projx}, \ref{projsum}, \ref{pxx} in Eq.~\ref{nxnx} gives,
\be
(n^x n^x)_{TT} = \frac{1}{2} \sin^2{(\psi + \theta)}\Big( \cos^2{\phi}(1+\cos{\psi}) - 1 \Big).
\label{epsnxnxtt}
\ee
In Sec.~\ref{sec:formalism} we discuss the setup where the observer is at $\infty$, \emph{i.e.,} $r = |\vec{x}|  \rightarrow \infty$, in that case the angle $\psi = 0$. Putting, $\psi = 0$ in Eq.~\ref{epsnxnxtt} results in,
\be
(n^x n^x)_{TT} = \frac{1}{2} (1 - \cos^2{\theta}) \Big( 2\cos^2{\phi} - 1 \Big).
\label{nxnxtt}
\ee

\section{An alternative proof of property \texorpdfstring{\eqref{eq:dotfourier}}{(2.16)}}

\label{appendix_fourier_derivative}

It is worth to clarify that in the usual treatment for the Fourier transform, property \eqref{eq:dotfourier} is usually proven by applying integration by parts and it follows after assuming that \be\lim_{t\to\pm\infty}g(t)\to 0\,.\ee However, in the case for gravitational memory, we can not make this assumption as by definition of memory,
 \be\lim_{t\to\pm\infty}h(t)\to h_{\rm final}\,.\ee
Here we want to provide a proof for the given property that does not rely on the vanishing of the boundary term, but only assumes the existence (finiteness) of the Fourier transform for the function under consideration. 
 
We want to compute $\Tilde{\dot{g}}(f)$. By using the rigorous definition of the derivative, we can write,
\be 
\Tilde{\dot{g}}(f)={\cal F}\left({\lim_{\d\to0}\frac{g(t+\d)-g(t)}{\d}}\right)
\ee
where for notation's convenience we have denote the Fourier transform as ${\cal F}$. Assuming the existence of the Fourier transform, and henceforth its inverse, we have,
\ba
\Tilde{\dot{g}}(f)&=&{\cal F}\left({\lim_{\d\to0}\int_{-\infty}^{\infty}\frac{e^{-2\pi i\d f'}-1}{\d}e^{-2\pi i f't}\Tilde{g}(f')df'}\right)\nonumber\\
&=&{\cal F}\left({\int_{-\infty}^{\infty}-2\pi i f'e^{-2\pi i f't}\Tilde{g}(f')df'}\right)\\
&=&\int_{-\infty}^{\infty} e^{2\pi i ft}\left({\int_{-\infty}^{\infty}-2\pi i f'e^{-2\pi i f't}\Tilde{g}(f')df'}\right)dt\,,\nonumber
\ea
In the last line we have used the definition of Fourier transform. Due again to the existence of the Fourier transform of $g(t)$ (and it's inverse), we can commute the integrations, obtaining,
\be
\Tilde{\dot{g}}(f)
 =\int_{-\infty}^{\infty} \d(f-f^\prime)(-2\pi) i f^\prime \Tilde{g}(f^\prime)df^\prime = -2\pi i f \Tilde{g}(f)\,.
\ee
We are unaware of this treatment in the literature.

\acknowledgments

We  acknowledge  funding  from the  National  Science  Foundation  grant  numbers  PHY-1613708 and PHY-2012195. MM was supported by the Fermi National Accelerator Laboratory (Fermilab) Award No. AWD00035045 during this work. CC was partly funded by the U.S. Department of Energy under grant number DE-SC0019470. We are grateful to Adam Burrows, Kei Kotake and David Vartanyan for allowing us to reproduce certain figures from their published works. We are grateful to Tanmay Vachaspati, George Zahariade and Michele Zanolin for fruitful discussions. We thank John Ellis, Marek Lewicki and Orlando L. G. Peres for useful feedback on the first version of the draft. 

\bibliography{refs}
\bibliographystyle{jhep}
\end{document}